\documentclass[onecolumn]{aastex61}
\usepackage{tcolorbox}
\usepackage{savesym}
\usepackage{soul}

\usepackage{amsmath}
\savesymbol{tablenum}
\usepackage{verbatim}
\usepackage{siunitx}
\restoresymbol{SIX}{tablenum}
\definecolor{orcidlogocol}{HTML}{A6CE39}
\usepackage{natbib}
\usepackage{hyperref}
\usepackage{graphicx}
\usepackage{float}
\usepackage{xcolor}
\colorlet{RED}{red}

\begin{document}
	\title{MOA-2010-BLG-328: Keck and HST Expose the Limits of Occams Razor in Microlensing}

	\correspondingauthor{Aikaterini Vandorou}
	\email{katievan@umd.edu}

	\author[0000-0002-9881-4760]{Aikaterini Vandorou}
	\affil{Laboratory for Exoplanets and Stellar Astrophysics, NASA/Goddard Space Flight Center, Greenbelt, MD 20771, USA}
	\affil{Department of Astronomy, University of Maryland, College Park, MD 20742, USA}
		\affil{Center for Research and Exploration in Space Science and Technology, NASA/GSFC, Greenbelt, MD 20771}

	\author[0000-0001-8043-8413]{David P. Bennett}
\affil{Laboratory for Exoplanets and Stellar Astrophysics, NASA/Goddard Space Flight Center, Greenbelt, MD 20771, USA}
\affil{Department of Astronomy, University of Maryland, College Park, MD 20742, USA}
\affil{Center for Research and Exploration in Space Science and Technology, NASA/GSFC, Greenbelt, MD 20771}

\author[0000-0003-0014-3354]{Jean-Philippe Beaulieu}
\affil{School of Natural Sciences, University of Tasmania,
	Private Bag 37 Hobart, Tasmania, 7001, Australia}
\affil{Sorbonne Universit\'e, CNRS, Institut d'Astrophysique de Paris, IAP, F-75014, Paris, France}
	
		\author{Aparna Bhattacharya}
		\affil{Laboratory for Exoplanets and Stellar Astrophysics, NASA/Goddard Space Flight Center, Greenbelt, MD 20771, USA}
		\affil{Department of Astronomy, University of Maryland, College Park, MD 20742, USA}
		\affil{Center for Research and Exploration in Space Science and Technology, NASA/GSFC, Greenbelt, MD 20771}
	
   \author[0000-0001-5860-1157]{Joshua~W.~Blackman}
	\affil{Physikalisches Institut, Universität Bern, Gesellschaftsstrasse 6, CH-3012 Bern, Switzerland}
	
		\author{Ian~A.~Bond}
		\affil{Institute of Natural and Mathematical Sciences, Massey University, Auckland 0745, New Zealand}
	
	\author[0000-0003-0303-3855]{Andrew A.~Cole}
	\affil{School of Natural Sciences, University of Tasmania,
		Private Bag 37 Hobart, Tasmania, 7001, Australia}
		
			\author[0000-0003-2302-9562]{Naoki Koshimoto}
\affil{Department of Earth and Space Science, Graduate School of Science, Osaka University, Tokyonaka, Osaka, 560-0043, Japan}
	
	\author[0000-0003-2388-4534]{Cl\'ement Ranc}
\affil{Sorbonne Universit\'e, CNRS, Institut d'Astrophysique de Paris, IAP, F-75014, Paris, France}

	\author[0000-0002-1530-4870]{Natalia E. Rektsini}
\affil{School of Natural Sciences, University of Tasmania,
	Private Bag 37 Hobart, Tasmania, 7001, Australia}
\affil{Sorbonne Universit\'e, CNRS, Institut d'Astrophysique de Paris, IAP, F-75014, Paris, France}

	\author[0000-0002-5029-3257]{Sean K. Terry}
\affil{Laboratory for Exoplanets and Stellar Astrophysics, NASA/Goddard Space Flight Center, Greenbelt, MD 20771, USA}
\affil{Department of Astronomy, University of Maryland, College Park, MD 20742, USA}
	\affil{Center for Research and Exploration in Space Science and Technology, NASA/GSFC, Greenbelt, MD 20771}

%	\author{Naoki  Koshimoto}
%	\affil{Okayama Astrophysical Observatory, National Astronomical Observatory of Japan, Asakuchi, Okayama 719-0232, Japan}
	
%	\author{Jean-Baptiste~Marquette}
%	\affil{Laboratoire d'astrophysique de Bordeaux, Univ. Bordeaux, CNRS, B18N, allée Geoffroy Saint-Hilaire, 33615 Pessac, France}
%	\affil{Sorbonne Universit\'es, UPMC Universit\'e Paris 6 et CNRS, 
	%	UMR 7095, Institut d'Astrophysique de Paris, 98 bis bd Arago,
	%	75014 Paris, France}

	\begin{abstract}
		
		We present high-resolution follow-up data of the microlensing event MOA-2010-BLG-328, using Keck and the Hubble Space Telescope. Keck data, taken 8 years after the event, reveal a strong lens detection enabling measurement of lens flux and source-lens relative proper motion. We find $\mu_{\rm rel, Hel} = 4.07 \pm 0.34\ \rm mas\ yr^{-1}$, with the lens $\sim10$ times fainter than the source. The lens was very faint in the Hubble passbands, and the small $\sim$35 mas lens-source separation made its detection difficult. However, we estimated lens magnitudes in Hubble bands by constraining its location to match the Keck $K$-band detection. \citet{Furusawa2013} reported a degenerate light curve, with viable models including microlensing parallax and lens orbital motion, or xallarap. We attempt to break this degeneracy by remodeling the event with constraints from high-resolution imaging, but find none of their models fit. Instead, fitting the follow-up data requires microlensing parallax, lens orbital motion, and a magnified binary companion to the host, with xallarap. The high-resolution data do break the “ecliptic degeneracy” common in Galactic bulge microlensing events with parallax signals. Models omitting any of these effects are excluded. However, a new degeneracy allows different combinations of parallax, orbital motion, xallarap, and source companion magnification to explain the light curve. These require either a nearby late M dwarf host or a more distant early M dwarf host, indistinguishable with NIR data alone.

	\end{abstract}
	

	\section{Introduction}
	
	There are many different types of techniques that can detect exoplanets (such as radial velocity and stellar transits), however, gravitational microlensing is currently the only one capable of finding low-mass planets in wide orbits, beyond the snow line. The direct imaging technique can also detect planets in wide-orbits, however, they are limited to more massive planets ($> 1\ M_{\rm jup}$). This wide-orbit parameter space is thought to be the dominant birth place of planets, as predicted by the core accretion theory \citep{Lissauer1993, Ida2004, Kennedy2006}. Gravitational microlensing cannot only detect `cold' planets down to Earth masses, but can also detect planets that orbit a range of different stars, such as M dwarfs (e.g. OGLE-2005-BLG-071, \cite{Udalski2005, Dong2009, Bennett2020}), and even white dwarfs (e.g. MOA-2010-BLG-477, \cite{Blackman2021Nature}). This is because the technique is not reliant on detecting light from the host star. A setback to this technique, however, is that the physical parameters of the system cannot be determined without the addition of higher order effects on the light-curve, or direct lens flux and proper motion measurements.
	
	Higher order effects on the light-curve can manifest as finite source effects, microlensing parallax, orbital motion of the lens, xallarap, and the magnification of a second source.  In the case of MOA-2010-BLG-328 (henceforth, MB10328) \cite{Furusawa2013} measured parallax, orbital motion and xallarap effects. They considered a ``parallax + orbital motion" model, which was degenerate between the close and wide solutions, and they considered a ``xallarap" only model. All these models were degenerate, yielding very different results. 
	
	 In cases such as these, high angular resolution follow-up observations \citep{Bennett2006, Bennett2007} with either ground based telescopes, like Keck Adaptive Optics, or space telescopes, like the Hubble Space Telescope (HST) can help confirm which solution is correct. Follow-up observations taken several years after the peak magnification can directly measure the lens-source relative proper motion in addition to the lens flux. Once these are added as constraints to the light-curve model, several mass-distance relations can be obtained, and thus the physical properties of the lensing system can be determined (e.g. \cite{Beaulieu2018, Bhattacharya2018, Bennett2020, Vandorou2020, Terrry2021, Terry2022}).
	
	In this paper we present high resolution adaptive optics (AO) observations from Keck-II's NIRC2 camera and the Hubble Space Telescope (HST) where we obtain constraints on the source-lens relative proper motion, and individual magnitudes, concluding that this is a binary lens binary source microlensing system. Using these constraints, we reanalyze the light-curve and find that incorporating additional higher-order effects is necessary beyond what was previously considered in \cite{Furusawa2013}. We therefore model the light-curve with finite-source effects, microlenisng parallax, orbital motion, xallarap, and magnification of the second source. This challenges the usual assumption that the ``simplest explanation is the best", a philosophical principle known as Occam's Razor. Our ``challenge" of it implies that this event presents a complex and degenerate set of solutions where multiple higher order effects must be considered. Keck and HST provide crucial follow-up data, but do not fully resolve the degeneracy, highlighting fundamental challenges in microlensing analysis.

	The paper is organised as follows; in Section \ref{section2} we present the microlensing event MB10328 with results from the detection paper by \cite{Furusawa2013} and the challenges they faced. In Section \ref{section3} we present new high resolution follow-up data that was acquired in 2018 from Keck and HST, and the analysis of this data. The results of this data are subsequently used as constraints for the light-curve remodeling, and presented in Section \ref{lightcurvesection}, where we also explore the various higher order effects and degeneracies. The physical parameters of the lens system are described in Section \ref{sec:planet system}. Finally, we discuss the overall results and conclude the paper in Section \ref{section6}. 

	This analysis is part of NASA's Keck Key Strategic Mission Support (KSMS) program, which has a primary goal of developing a mass-measurement method \citep{BennettWFIRST} for the Nancy Grace Roman Space Telescope (formerly, WFIRST) \citep{Spergel2015}. This program will be dedicated to a microlensing survey of exoplanets towards the Galactic Bulge. The findings of this program will complement other statistical studies, such as those of transiting planets \citep{Borucki2011}.

	%mention parallax stuff 
	
	\section{Microlensing event MOA-2010-BLG-328}
	\label{section2}

The microlensing event MB10328 was first detected by the Microlensing Observations in Astrophysics (MOA) group on 16 June 2010. The event is located at (R.A., Dec) = (17:57:59.12, --30:42:54.63) and Galactic coordinates $(l, b) = (-0.16, -3.21)$ degrees. An anomaly in the predicted single lens light-curve model was observed, and an alert was issued to other microlensing groups on 27 July 2010. Shortly after other follow-up groups began to observe the event, and data was obtained from the Microlensing Follow-Up Network ($\mu$Fun; \cite{Gould2006}), the Probing Lensing Anomalies NETwork (PLANET; \cite{Beaulieu2006}), Microlensing Network for the Detection of Small Terrestrial Exoplanets (MiNDSTEp; \cite{Dominik2010}) and RoboNet \citep{Tsapras2009}.

 The data, modelling and analyses are presented in \cite{Furusawa2013}. Their light-curve model revealed the effects of a microlensing parallax due to the orbital motion of the Earth \citep{Smith2003} and the orbital motion of the source star, which is often referred to as the xallarap effect. Although they do not completely exclude the xallarap solution, the data prefers the parallax plus orbital motion solution by $\Delta \chi^2 =5$. At such a low level of significance this could be influenced by systematics, which often affects parallax solutions in microlensing events. For events that have a relatively weak microlensing parallax, approximate symmetry exists in the plane where the lens is replaced by its mirror image. That symmetry is broken for this event however, and the $u_0 < 0$ and $u_0 > 0$ solutions differ by $\Delta \chi^2 = 78$, where $u_{0}$ is the impact parameter in units of Einstein radii.
 
The source color and magnitude were derived from CTIO V and $I$ band photometry, and the observed I band magnitude of the source star from the model and OGLE data. In combination with the color-color relations from \cite{Bessell1988} they found the $V$, $I$ and $K$ magnitudes of the source for the parallax plus orbital motion solutions and the xallarap solution. These are presented in Table \ref{furusawasource}.

 \setlength{\tabcolsep}{10pt}
\begin{deluxetable}{ l  c  c  c}
	\tablecaption{\cite{Furusawa2013}'s source magnitude estimates from their accepted best fit models. \label{furusawasource} }
	\tablewidth{0pt}
	\tablehead{
		&  \multicolumn{2}{c}{Parallax plus orbital motion } & \\
		\colhead{Magnitude}	 &   ($u_0 < 0$)  &($u_0 > 0$) & \colhead{Xallarap} 
	}  % end header.
	%\hline
	%\hline
	\startdata
	$I_S$& 19.49 $\pm$ 0.04 & 19.49 $\pm$ 0.04 & 19.46 $\pm$ 0.04  \\   
	$V_S$&19.93 $\pm$ 0.11 & 19.93 $\pm$ 0.11 &   19.90 $\pm$ 0.11\\
	$K_S$&$16.94\pm 0.26$ & $16.97 \pm 0.26$ & 17.12 $\pm$ 0.27 \\
	\enddata
\end{deluxetable}	

From these source colors and magnitudes they calculated the angular Einstein ring radius, $\theta_E$, and relative source-lens proper motion, $\mu$. Combining these with a measured parallax, they obtained physical parameters for the system for the close and wide parallax plus orbital motion model, assuming a source distance $D_{\rm S} = 8.0 \pm 0.3$. They found a system with lens mass of $M_{\rm L}  \sim 0.1\ M_{\odot}$ with a super Earth planetary companion. For the xallarap model, the probability distributions of the physical parameters were estimated using a Bayesian analysis. These results indicated a  $M_{\rm L}  \sim 0.6\ M_{\odot}$ lens mass with a Saturn-mass planetary companion.

\begin{figure*}[t!]
	\centering
	\includegraphics[width=18cm]{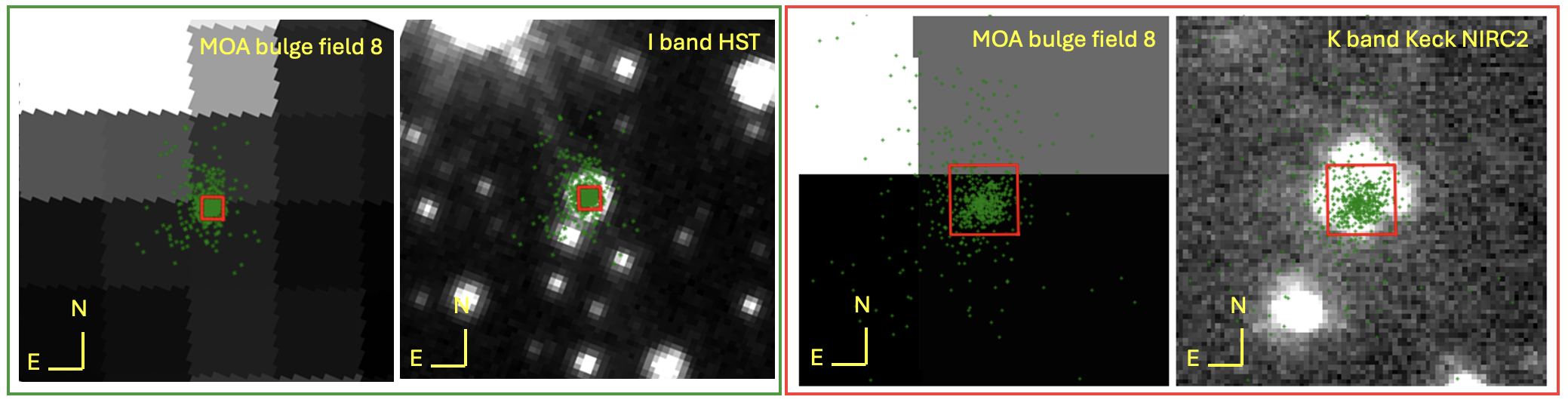}
	\caption{Target star identification in Keck and HST using MOA survey data. Target is identified by individual centroid positions in green, with red box indicating the 1$\sigma$ error of the scatter. }
	\label{fig:moa-id}
\end{figure*}

MB10328 was reobserved with Keck and HST, and the target was identified in both using the initial MOA survey data as seen in Figure \ref{fig:moa-id}. A sample of 624 difference images were used where the event was undergoind lensing, and the individual centroid positions on this sample were measured with the optomised PSF used for the light-curve photometry. The scatter of this measurement determined the error box boundaries in the MOA frame, with the width being $\sim 0.1$ arcseconds. The MOA frame was matched to the HST frame using an affine transorm relation, derived by cross referencing the same stars in both frames.  This resulted in a total of 44 matching star-pair positions, which were used to derive the transformation coefficients. The MOA-HST transformation was then used to transform the MOA error box and the clustering of individual measurements around the star. The MOA and Keck frames were matched in a similar process, finding 50 matching star-pairs. Transforming the MOA measurements onto the Keck frame resulted in the same star being identified as in the HST images.

Using high resolution follow-up observations with Keck and HST we directly measure the source and lens flux, in addition to the relative proper motion (direction and magnitude). With this data as constraints we attempt to break the degenaracies in the light-curve model and find a plausible solution for this system.

	\section{High Resolution Follow-up Observations with Keck and HST}
	\label{section3}
	
	\subsection{Keck NIRC2}
	High angular resolution follow up observations of MB10328 were conducted in May and August 2018 using the NIRC2 laser guide star adaptive optics (AO) system on the Keck Telescope in Hawaii.
	
\begin{figure*}[t!]
	\centering
	\includegraphics[width=15cm]{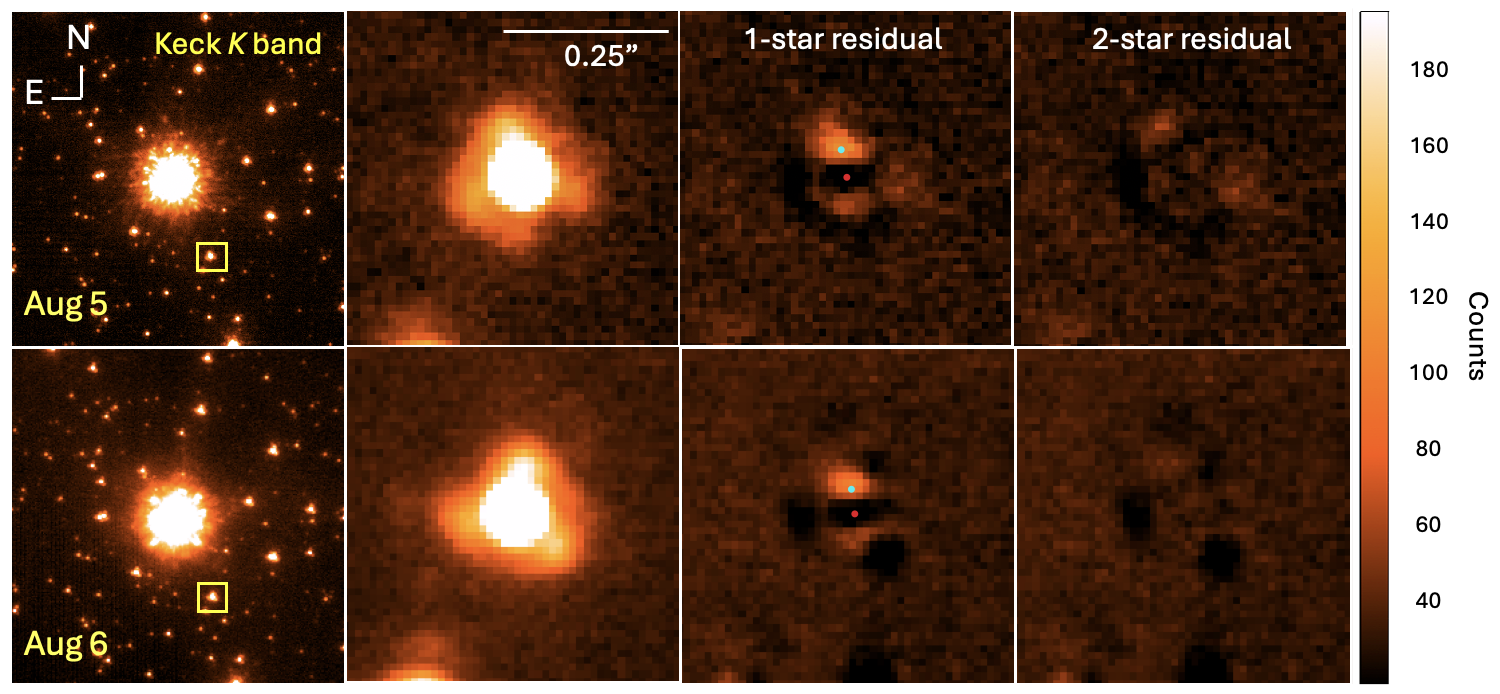}
	\caption{Keck $K$ band data from August 5 (top panels) and August 6 (bottom panels), showing a zoom of the target, the 1-star residual where the source is subtracted, and the 2-star residual where both the source and lens are subtracted. In the 1-star residual frame the source and lens positions are indicated in red and cyan, respectively. }
	\label{fig:328residuals}
\end{figure*}

	On May 26 2018, MB10328 was observed using Keck II's NIRC2 instrument with the wide camera which has a plate scale of 39.69 mas/pixel. The images were taken in the $K$ short band ($\lambda_c=2.146\ \mu m$, $K_s$, hereafter $K$). Eight images were used, with an exposure time of 30 seconds and a dither of $\sim 2 '' $.  The images were reduced by correcting for the dark current and flat-field (e.g. \cite{Beaulieu2016, Batista2014}). The images were also sky-corrected. An astrometric solution was found by cross-identifying 76 stars with our reprocessing of the VVV catalogue (VISTA Variables in the Via Lactea) \citep{Minniti2010}, and the frames were stacked using SWarp \citep{Bertin2010}. The stellar fluxes were measured using aperture photometry with SExtractor \citep{Bertin1996}. Thus we measured a $K$ band magnitude for the source and lens to be:
	\begin{equation*}
	K_{\rm SL} = 16.08 \pm 0.09
	\end{equation*}
On August 5 and 6 2018, MB10328 was observed again, using Keck II's NIRC2 narrow camera which has a plate scale of 9.94 mas/pixel. The images had an exposure time of 30 seconds in the $K$ band. 10 images were obtained on August 5, and 30 on August 6. The images were reduced using `KAI', which is a data-reduction pipeline specifically designed to reduce Keck AO data \citep{Lyke2017, Lockhart_2019}. The pipeline includes the standard dark current correction, flat-fielding and sky correction, in addition to bad pixel and cosmic ray masking. The images are combined using a Jackknife routine (\citet{Quenouille1949, Quenouille1956, Tukey1958}) where $\rm N-1$ frames are stacked, where N is the number of total frames. Therefore, we produce 10 different stacks of 9 frames for the August 5 observations, and 30 different stacks of 29 frames for the August 6 data, since one frame is removed each time. The target star can be seen in Figure \ref{fig:328residuals} for both nights. This process was repeated for both August 5 and 6, and the subsequent analyses on the images was conducted separately, and for the final result we use the average of these two sets of data (see Table \ref{mcmc}). Using this Jackknife stacking method is standard procedure now when analysing high resolution follow-up images with Keck, with further details to be found in \cite{Terrry2021, Bhattacharya2021}. 

 The quality of the August 6 data was good with a Strehl ratio and FWHM across the images with an average of 0.315 and 58 mas, respectively. On the other hand, the Strehl ratio and FWHM for the August 5 data was slightly worse at 0.226 and  71 mas, respectively. The lens, however, was robustly detected on both nights.  
 
 %{\color{red} Should I take the average or should i just take the best chi2? Discuss more about changes in images eg from Sean's paper.}

\subsubsection{PSF Fitting Photometry $\&$ Astrometry}	

Since the predicted separation between the source and lens is estimated to be about 30-40 mas in 2018, we can analyse the Keck data to partially resolve the two stars, therefore measuring their relative proper-motion (magnitude and direction) and their individual fluxes.  

We use a modified version of \textit{DAOPHOT-II} \citep{Stetson1987} called \textit{DAOPHOT-MCMC} \citep{Terrry2021} which runs Markov Chain Monte Carlo (MCMC) sampling on the target star. The first step to this routine is to build an empirical PSF model using reference stars of similar brightness and PSF shape in the target's field. This is done for each individual jackknife frame, as the PSF shape can vary from image to image. We select reference stars with magnitude ranges of $ -0.3 < m < 0.3$ and separations $-3'' < r <  3''$ from the target.  

\begin{figure*}[t!]
 	\centering
 	\includegraphics[width=14cm]{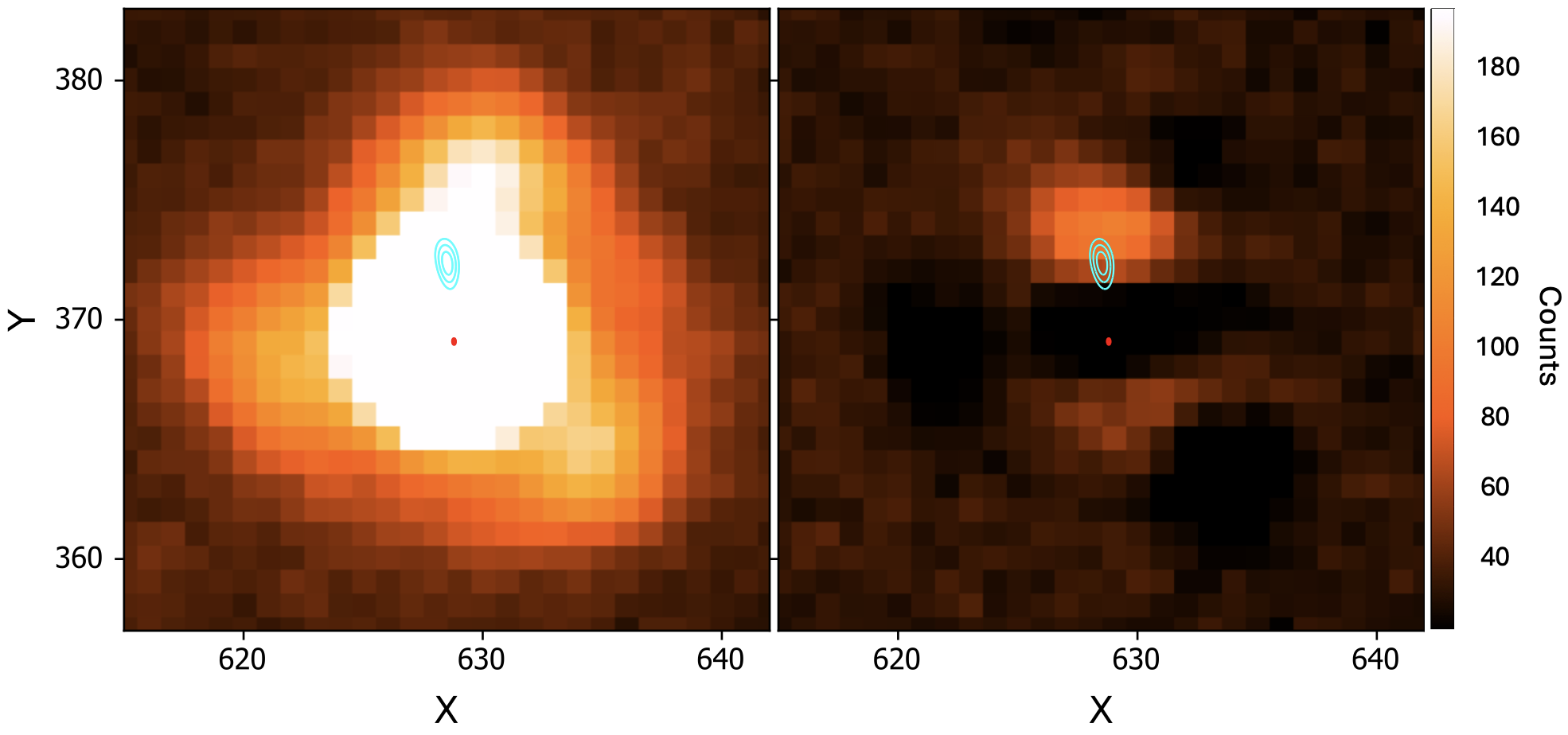}
 	\caption{Best fit MCMC contours (68.3\%, 99.5\%, 99.7\%) of the X and Y pixel positions of the source (red) and lens (cyan). The contours are overplotted on the stacked Keck Narrow image (left) and the 1-star residual image (right).}
 	\label{fig:328contors}
 \end{figure*}

The PSF model is fit to the target which produces a single-star residual. A clear feature is seen North of the target in both the residual images from August 5 and 6, as shown in Figure \ref{fig:328residuals}. We then fit a 2-star PSF model (simultaneously fit two stellar PSF models) to the target star in our Keck image, which produced a featureless residual as also seen in Figure \ref{fig:328residuals}. Running the MCMC routine with the 2-star PSF model produces a singificantly better the fit compared to the 1-star fit with $\Delta\chi^2 \sim 350$.  We plot the best fit MCMC contours on Figure \ref{fig:328contors} of the pixel positions of the source (red) and lens (cyan). The position of the lens star does not match the peak of the residual feature because some of the flux of the lens is included in 1-star profile.

The main source of uncertainty in our data stems from the PSF variations between the individual Keck frames, and across each frame. Using the jackknife method to stack our frames allows us to determine this uncertainty between the images. The errors are estimated by the $\sqrt{N-1}$ multiplied by the root-mean-square of the best fit parameters from each $N-1$ combinations as shown in the equation below:

\begin{equation*}
	SE(x) = \sqrt{\frac{N-1}{N} \sum(x_i - \overline{x})^2},
\end{equation*}

 where N is the number of stacked images, $x_i$ is the parameter measured in each of the stacked images, and $\overline{x}$ is the mean of $x$ for all N stacked images. Further details on the application of the jackknife method on high-resolution follow-up data can be found in \citet{Bhattacharya2018, Terry2022}.

 \subsubsection{Unusual PSF shapes}

 It is easy to notice the unusual PSF shape of the target star in the Keck images on both nights, but especially on August 6. This is not surprising for ground based telescopes, even with AO correction, and can be caused by a number of reasons such as issues with the AO system, optical aberrations, or atmospheric effects. This is why we use the jackknife stacking method, and build an empirical stellar PSF model using single stars close to the target. 
 
\begin{figure*}[h!]
 	\centering
 	\includegraphics[width=8cm]{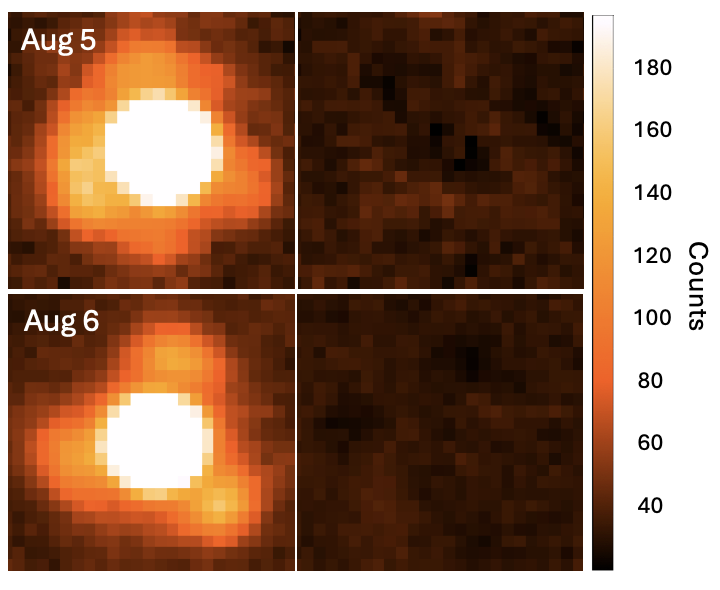}
 	\caption{Example of a PSF star and its residual from each night of observation that is used to build an empirical PSF model.}
 	\label{fig:328psfstars}
 \end{figure*}

The stars across the frame have similar PSF shapes to that of the target, therefore, these unusual features are not unique to the target star, and when the PSF model is subtracted (as seen in the 1 and 2-star residual frames of Figure \ref{fig:328residuals}) these features disappear. In Figure \ref{fig:328psfstars} we show an example of a PSF star used to build an empirical PSF model for each night of observation. We see similar shapes to that of the target, and the smooth residuals prodiced when the PSF model is subtracted.
 
 \subsubsection{Source-lens relative proper motion and flux ratio}

 \setlength{\tabcolsep}{10pt}
 \begin{deluxetable}{ l  c   }
 	\tablecaption{The separation, source-lens relative proper motion and flux ratio average from the 5th and 6th of August 2018 Keck data. \label{mcmc} }
 	\tablewidth{0pt}
 	\tablehead{
 		\colhead{Parameter} 	&   \colhead{Average} 
 	}  % end header.
 	%\hline
 	%\hline
 	\startdata
 	Separation (mas)& $34.885\pm0.845$  \\   
 	$\mu_{\rm rel, Hel, East}$ ($\rm mas\ yr^{-1}$)&$0.708\pm0.103$  \\
 	$\mu_{\rm rel, Hel, North}$ ($\rm mas\ yr^{-1}$)&$4.371\pm0.104$   \\
 	Lens / Source flux &$0.119 \pm0.003$    \\
 	%$\chi^2$ & 486 \\
 	%$dof$& 477 &  454\\
 	\enddata
 \end{deluxetable}

 The Keck follow-up observations in August 2018 where taken 7.93 years after the peak magnification in 2010. In \cite{Furusawa2013} they estimated three different source-lens proper motions, two for the close and wide solution of the parallax plus orbital motion model ($\mu_{\rm close}=5.71 \pm 0.70$ $\rm mas\ yr^{-1}$, $\mu_{\rm wide}=4.72 \pm 0.79$ $\rm mas\ yr^{-1}$), and one for their xallarap model ($\mu_{\rm xal}=4.03 \pm 0.26$ $\rm mas\ yr^{-1}$). 
 
 By directly measuring the separation between the source and lens positions in the Keck data, and knowing that 7.93 years have passed since peak magnification, we calculate the heliocentric relative proper motion to be $(\mu_{\rm rel, Hel, E}, \mu_{\rm rel, Hel,N})=(0.708, 4.371) \pm (0.103, 0.104) \ \rm mas\ yr^{-1}$, where E and N denote the East and North direction, respectively. Therefore the amplitude of the relative proper motion vector is $\mu_{\rm rel,1Hel} = 4.428 \pm 0.104\ \rm mas\ yr^{-1}$. 
 
 By analysing the PSF residual we can also calculate the flux distribution of both source and lens star. We find a best fit flux ratio of $f_{\rm lens}/f_{\rm source} = 0.119 \pm 0.003$. Since we know the flux contribution of each star and the total blended flux from the wide Keck image, we can derive the individual K-band magnitudes for the lens and source. To do this we use the following set of equation, and solve them simultaneously: 
 \begin{equation}
 	K_{\rm L} - K_{\rm S} = -2.5\log_{10}(f_{\rm L}/f_{\rm S})
 	\label{sim1}
 \end{equation}
 \begin{equation}
 	\label{sim2}
 	K_{\rm SL}= -2.5\log_{10}(10^{-0.4K_{\rm L}}+10^{-0.4K_{\rm S}})
 \end{equation}
 
 The L and S notations represent the lens and source respectively. $K_{\rm SL}$ is the K band magnitude of the total flux, which is measured from the Keck wide image ($K_{\rm SL} = 16.08 \pm 0.09$). Therefore, using equations \ref{sim1} and \ref{sim2} we calculated a lens and source magnitude in K band to be: 
 
 \begin{gather*}
 	K_{\rm L} = 18.513 \pm 0.094\ \rm mag\\
 	K_{\rm  S} = 16.202 \pm 0.090\ \rm mag.
 \end{gather*}
 
We can compare this source $K$ band value with those found by \citet{Furusawa2013} (see Table \ref{furusawasource}), and we find that our measured Keck value is $\sim0.8$ mag brighter. This is unsurprising since the system likely has a binary source. Therefore, when remodeling the light-curve in Section \ref{lightcurvesection}, a binary lens binary source scenario is explored which includes the higher order effects of parallax, orbital motion and xallarap.

\subsection{HST WFC3/UVIS}
	
The microlensing event  MB10328 was observed twice with the Hubble Space Telescope's (HST) WFC3/UVIS camera. The first set of observations were taken 27 May 2018 in the F814W and F555W filters as part of the GO-15455 proposal (PI: Bennett).  \footnote{The HST data presented in this article were obtained from the Mikulski Archive for Space Telescopes (MAST) at the Space Telescope Science Institute. The specific observations analyzed can be accessed via \dataset[doi: 10.17909/yftx-fp58]{https://doi.org/10.17909/yftx-fp58}.}

All data were flat-fielded, stacked and corrected for disortions. Pixel area distortion, astrometry and PSF photometry were carried out on the F814W and F555W data separately with a modified version of the code used by \cite{Bennett2015} and \cite{Bhattacharya2018}. This code does not use any resampling, and instead analyses the data from the original images in order to mitigate any loss in resolution that a combined, dithered, undersampled image would. In addition, due to HST's relative smaller aperture and large undersampled pixels, its images have worse angular resolution compared to Keck. However, HST compensates for this by having a much more stable PSF.

For the HST 2018 data, 15 $\times$ 84 sec dithered exposures were taken in the F814W filter ($I$-band) and 16 $\times$ 91 sec dithered exposures were taken in the F555W filter ($V$-band). Since the 2018 HST data are only taken 3 months prior to the narrow Keck images, we expect the separation to be approximately the same (i.e. $\sim$35 mas). The coordinate transformation between the Keck and HST images was done using 16 bright stars, yielding: 
\begin{gather*}
	x_{\rm HST, 2018} = -0.23343\ x_{\rm Keck} + 0.09157\ y_{\rm Keck} + 630.74321 \\
	y_{\rm HST, 2018} = -0.09162\ x_{\rm Keck} - 0.23335\ y_{\rm Keck} + 779.36899.
\end{gather*}

The avergage RMS scatter for these relations is $\sigma_x \sim 0.02$ and $\sigma_y \sim 0.02$.

We fit a single star PSF model to produce the 1-star residual (see Figure \ref{fig:328hst}), and then a two star PSF model. The single star residual model does not visually show the signal that we expect from two blended stars (as seen in the Keck images). However, from the Keck images we have a clear result on the position of the lens relative to the source. Even though we cannot ``see by eye" a residual signal, we run a constrained fit using an MCMC routine on the position and flux of the lens and source, using the constraints on $\mu_{\rm rel}$ from the Keck 2018 data. We calibrate the HST data to the OGLE-III catalog \citep{Szymanski2011} by matching bright isolated stars. Therefore we measure the V and I magnitudes of the source (primary and secondary source total) and the lens which has a known position relative to the source. These values are shown in Table \ref{hst}. For the combined total flux of the blended target (lens and source) in HST we find:

 	\begin{figure*}[t!]
	\centering
	\includegraphics[width=17cm]{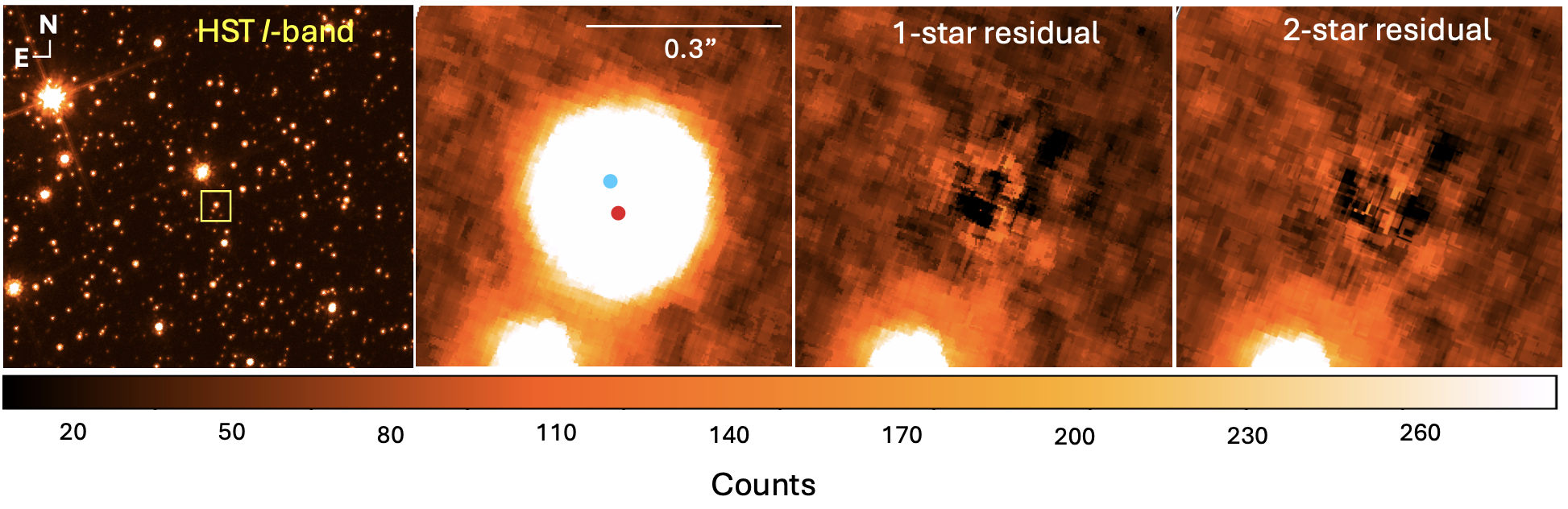}
	\caption{From left to right: first panel shows the stacked image of the 2018 HST $I$-band data, the second shows a zoom of the target with the constrained lens and source positions indicated in cyan and red, respectively. The third and fourth panels show the 1 and 2 star fits to the target, respectively. The faint lens star cannot be directly identified in these images, however, since we know the lens-source separation from the Keck analysis, we can measure the lens and source magnitudes with a fit that constrains their separation to be the same as measured from the Keck data.}
	\label{fig:328hst}
\end{figure*}

\begin{gather*}
	(V-I, I)_{\rm HST,total} = (2.108, 19.275) \pm (0.081, 0.050).
\end{gather*}

The 1-star model yields a $\chi^2 = 510.6262$ compared to that of the 2-star model with $\chi^2 = 510.8950$. Although the 1-star model yields a higher $\chi^2$, the difference is very small, suggesting that the results of the 2-star model can also be reasonable. Nevertheless, going forward we only use the total color and magnitude of the source and lens as a constraint in the image constrained light-curve model (see Section \ref{lightcurvesection}). The lens is faint in the $V$ and $I$ passbands, and the lens-source separation is small compared to the HST FWHM. Therefore, getting a good measurement of both the lens-source flux ratio and their separation is difficult. However, if the lens-source separation is known, then the lens flux can be measured in a constrained fit \cite{Bennett2007}. This constrained fit yields a measurement of the lens $I$ band magnitude and a very marginal measurement of the $V$ band magnitude.

\setlength{\tabcolsep}{10pt}
\begin{deluxetable}{ l  c  c  }
	\tablecaption{The lens/source flux ratio (where ``source" includes the primary and secondary source), and the individual calibrated magnitudes in I and V from the HST 2018 data. \label{hst} }
	\tablewidth{0pt}
	\tablehead{
		\colhead{Parameter} 	&   \colhead{$I$} &   \colhead{$V$}  
	}  % end header.
	%\hline
	%\hline
	\startdata
	 Lens/Source flux& 0.048 $\pm$ 0.013  &   0.024 $\pm$ 0.015\\   
	Lens (mag) & 22.628 $\pm$ 0.286 & 25.411 $\pm$ 0.651 \\
	Source (mag)  & 19.326 $\pm$ 0.052 &  21.410 $\pm$ 0.049 \\
	\enddata
\end{deluxetable}

	\section{Image Constrained Light-Curve Model}
	\label{lightcurvesection}
	
	%$I_{\rm S}$ & 17.66$\pm$ 0.04 & 17.66$\pm$ 0.04& 17.63 $\pm$ 0.04 & 19.609 $\pm$ 0.023\\
	%$(V-I)_{\rm S}$ &0.70 $\pm$ 0.10 &0.70 $\pm$ 0.10& 0.70 $\pm$ 0.10& 2.189 $\pm$ 0.041\\
	%$I_{\rm S2}$ &-- & --& --& 20.864 $\pm$ 0.131\\
	%$(V-I)_{\rm S2}$ &-- & --&-- & 1.918 $\pm$ 0.373\\
	%$\theta_{\rm E}$(mas) & 0.98 $\pm$ 0.12 & 0.83 $\pm$ 0.14 &$ 0.68\pm0.04$ & 0.629 $\pm$ 0.023\\ 

To obtain a full solution of the lens system's physical properties (lens mass, distance, planet mass, and projected separation), a combination of light-curve model and high resolution follow-up parameters is needed. This process requires that we first remodel the light-curve using constraints on $\mu_{\rm rel}$, the lens and source flux, and color from the high resolution data as described in \cite{Bennett2024}. The MCMC results from this modeling procedure are therefore consistent with the high resolution follow-up data. Models that are consistent with the light-curve data, but not the high angular resolution follow-up data are excluded by the large $\chi^2$ contributions from the high resolution data constraints. Then, the lens system parameters can be determined by summing over the light-curve model MCMC results using a Galactic prior, in combination with mass-distance and mass-luminosity relations. We discuss these relations in Section \ref{sec:planet system}.

With the measured relative proper motion magntitude and direction, $\mu_{\rm rel}$ from Keck and its uncertainty, we can constrain the Einstein ring radius $\theta_{\rm E}$, and the direction of the microlensing parallax $\pi_{\rm E}$, because the vector $\pi_{\rm E}$ is parallel to the $\mu_{\rm rel}$ vector once it has been converted from heliocentric to geocentric reference frame (see Section \ref{sec:planet system} and equations therein). We also constrain the lens star brightness using the measured lens $K$ flux from Keck, and the combined source and lens $I$ and $V$ flux from our HST measurements.

\begin{figure*}[t!]
	\centering
	\includegraphics[width=11cm]{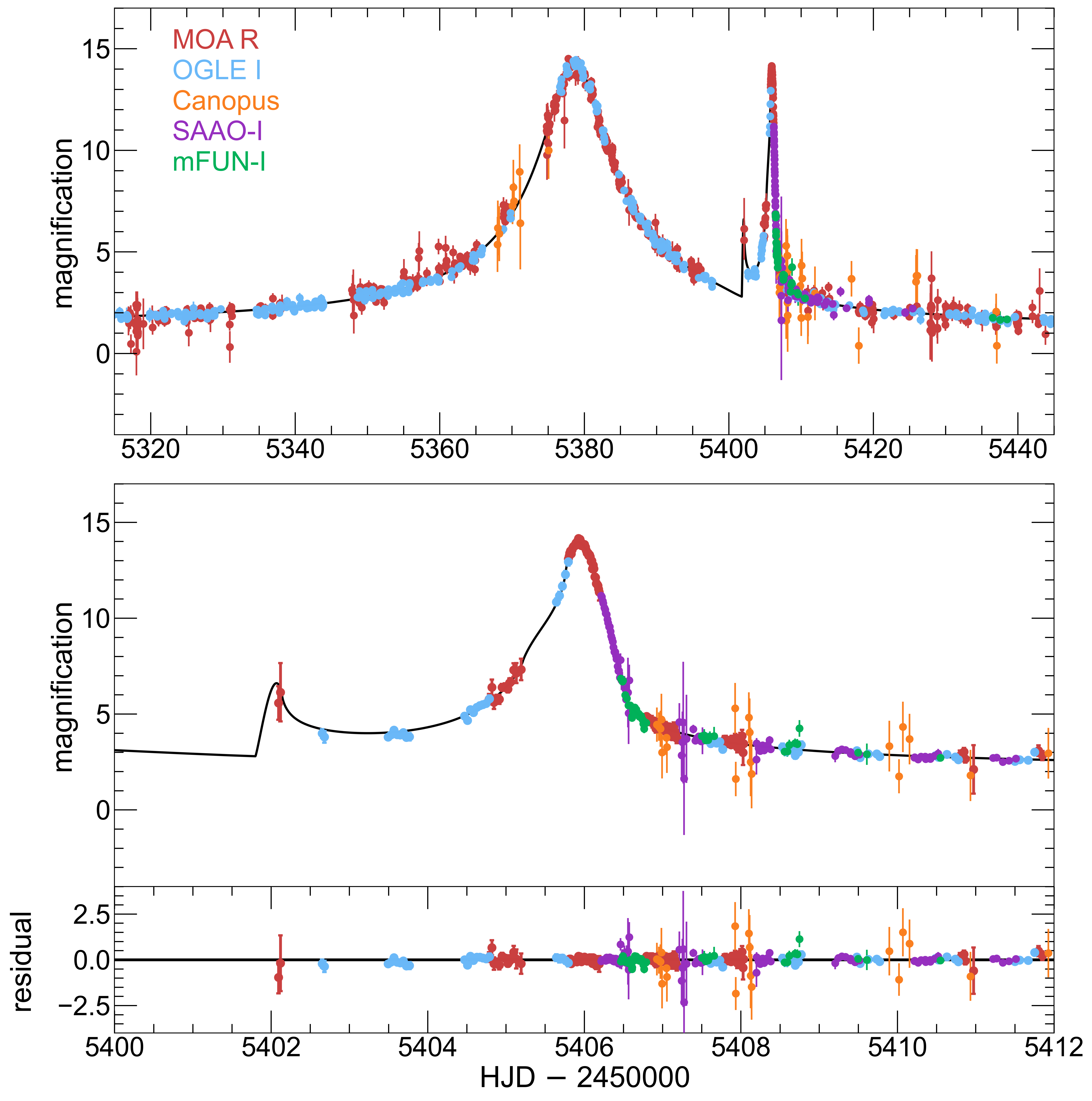}
	\caption{Best fit 2L2S light-curve for MB10328 using constraints from the Keck and HST high resolution follow-up data. The middle panel is a close-up of the upper pannel, indicating the planetary anomally. The lower plot shows the residuals. The parameters are described in Section \ref{lightcurvesection} and Table \ref{tab-mparams2}.}
	\label{fig:lightcurve}
\end{figure*}

%\subsection{\textbf{Modeling with Higher Order Effects}}
%\label{sec:modeling and higher order effects}

The modeling for MB10328 is particularly challenging as it includes many higher order effects, which are not typically modeled simultaneously as seen in \cite{Furusawa2013}. However, in the remodeling of this event we include all higher order effects; microlensing parallax, orbital motion, xallarap and magnification of the second source. The modeling is done with a modified version of the light-curve modeling code {\fontfamily{lmtt}\selectfont eesunhong}\footnote{\url{https://github.com/golmschenk/eesunhong}} \citep{Bennett1996, Bennett2010}. The {\fontfamily{lmtt}\selectfont eesunhong} code follows an image-centered ray shooting method and includes constraints on both the brightness and relative proper motion of the source and lens from the high resolution follow-up data. For a full discreption of this methodology we refer the reader to \citet{Bennett2024}.  To speed up the light-curve modeling, most of the modeling was done with photometry data that was binned to an interval of 0.01 days, after verifying that this binning had no significant effect on the model parameters. The light-curve is plotted in Figure \ref{fig:lightcurve}. It does not include the SAAO and $\mu$Fun V-band data due to the two sources not being the same color. 

In Table \ref{tab-mparams2}, we show the light-curve model parameters with the values and uncertainties for the Einstein radius crossing time ($t_{\rm E}$), time of closest approach ($t_0$), source crossing time ($t_{*}$), impact parameter ($u_0$), the instantaneous projected separation which is scaled to $\theta_{\rm E}$ ($s$), the host-planet mass ratio ($q$), and the angle between the source trajectory and traverse line that passes through the two lenses ($\alpha$). The model parameters that are essential for obtaining solutions to the lens system's physical parameters are  $q$, $t_{\rm E}$, $t_*$, and the microlensing parallax $\pi_{\rm E}$. 
	
Another essential parameter is the source distance $D_{\rm S}$, which is needed to determine the mass of the lens from the relative proper-motion constraint, as shown in equation \ref{eq:pirel}. Therefore, we include $D_{\rm S}$ as a light-curve model parameter and use the \citet{Koshimoto2021a} Galactic model as a prior.

The Galactic bulge is located close to the ecliptic plane therefore microlensing events with significant microlensing parallax signals often exhibit an ``ecliptic degeneracy" \citep{Pointdexter2005} which involves replacing the microlensing geometry with its mirror image. In the case of this event, the directions of relative proper motion, $\mu_{\rm rel}$, differs by more than $160^\circ$, so the Keck $\mu_{\rm rel}$ measurement removes this degeneracy. 

Nevertheless, an examination of the light-curve model analysis of this event indicated a significant model degeneracy involving microlensing parallax and the orbital motions of both the binary source star system and the lens star plus planet system. These degeneracies are related to the continuous degeneracy between constraints based on angular Einstein radius, $\theta_{\rm E}$, and host star $K$ band magnitudes for M dwarf host stars located at $\sim 4 $kpc as indicated in Figure 7 of \cite{Fukui2015}, Figure 5 of \cite{Naoki2020} and Figure 7 of \cite{Terry2024}. These constraints overlap for M dwarf masses spanning a range of$\sim 5$ in mass. This sort of degeneracy can often be resolved with a measurement of the microlensing parallax \citep{Bhattacharya2018, Bennett2020, Terry2024} but in this case, the planetary orbital motion, magnification of the 2nd source, and the conversion from the Heliocentric to Geocentric relative proper motion conspire to retain an approximate continuous degeneracy. The
Markov Chain revealed a $\chi^2 $ 2 barrier that separated this degeneracy into two largely isolated regions of large and small $|\pi_{\rm E}|$. This gives the bimodal distributions for $\pi_{\rm E,N}$, $\pi_{\rm E,E}$, and the star-planet angular velocity in the plane of the sky, $\omega$, as seen in Figure  \ref{fig:328corner}.

%For a number of events, such as OGLE-2005-BLG-071 \citep{Bennett2018}, MOA-2007-BLG-192 \citep{Terry2024}, and OGLE-2012-BLG-0950 \citep{Bhattacharya2018}, the microlensing parallax degeneracy is resolved primarily in the $\pi_{\rm E, N}$ component, and enables precise determination of the 2-dimensional $\pi_{\rm E}$ vector. However, for this event, the orbital motions of the binary source star system and the host star plus planet lens system conspire to maintain such degeneracy. 

To investigate this degeneracy and understand how the light-curve model parameters are being affected, we make a diagonal cut in the ($\pi_{\rm E,E}, \pi_{\rm E,N}$) space on the corner plot (Figure \ref{fig:328corner}), at $\pi_{\rm E,E} = \pi_{\rm E,N} - 0.33$.  In Table \ref{tab-mparams2} we present the light-curve model parameters associated with the ``large" and ``small" parallax values, where ``large" refers to $\pi_{\rm E, N} > \pi_{\rm E,E}  - 0.33$, and ``small" refers to  $\pi_{\rm E, N} <  \pi_{\rm E,E}  - 0.33$. We also do this in Table \ref{tab:params} while presenting the lens system's physical parameters. In Section \ref{deg} we discuss this degeneracy further.

%\subsection{Applying Constraints from High Angular Resolution Follow-up Observations on Light-curve Models}

% talk about mu_rel constraint 
% can be converted to mu_rel_Geo 
% tE and thetaE

%lens and source brightness constraint 
% Total 

\subsection{Source Color and CMD}

Using the measured total magnitude and color from HST as a constraint on the light-curve model, we find MCMC distributions of the primary source's (S1) magnitude and color to be:  
\begin{gather*}
	(V-I, I)_{\rm S1} = (2.188 , 19.613) \pm (0.038, 0.021)\ \ \ \ {\rm for\ small}\ \pi_{\rm E}, \\
	(V-I, I)_{\rm S1}  = (2.190 , 19.617) \pm ( 0.030, 0.017)\ \ \ \ {\rm for\ large}\ \pi_{\rm E},
\end{gather*}
and for the secondary source (S2): 
\begin{gather*}
	(V-I, I)_{\rm S2} = (2.036 , 20.807) \pm (0.502, 0.244)\ \ \ \ {\rm for\ small}\ \pi_{\rm E}, \\
	(V-I, I)_{\rm S2}  = (1.894 , 20.846) \pm (0.324, 0.183)\ \ \ \ {\rm for\ large}\ \pi_{\rm E}.
\end{gather*}

To calculate the source angular size of the primary source, we need the dereddened magnitude and color. The extinction to the source star is determined by using the red clumb stars within 120" of the MB10328 event from the OGLE-III catalogue \citep{Szymanski2011}. Therefore, we find $A_I = 1.817$ and $A_V = 3.247$. Using the following relation from \cite{Boyajian_2014}: 

\begin{equation}
	\label{eq:surface brightness}
	\log_{10}[2\theta_{*}/(1\rm mas)] = 0.5014 + 0.4197(V-I)_{\rm s0} - 0.2I_{\rm s0},
\end{equation}

we calculate the primary source's angular radius $\theta_{*} = 0.910 \pm 0.042\ \mu \rm as$ (for small $\pi_{\rm E}$) and $\theta_{*} = 0.911 \pm 0.032\ \mu \rm as$ (for large $\pi_{\rm E}$ ). We can estimate the relative source-proper motion using $\mu_{\rm rel, G} = \theta_{*}/t_* = 3.587 \pm 0.436$ mas $\rm yr^{-1}$ (for small $\pi_{\rm E}$) and $\mu_{\rm rel, G} = 5.099 \pm 0.243$ mas $\rm yr^{-1}$ (for large $\pi_{\rm E}$) , where the notation $G$ represents the geocentric reference frame. This cannot be directly compared to the relative proper motion measured in Keck, since Keck is in the heliocentric reference frame. In Table \ref{tab:params} we show our converted Keck proper motion, after converting it to the geocentric reference frame (see Section \ref{sec:planet system} for details).

From the light-curve model we also find the MCMC distributions of the lens $V$ and $I$ measurements:

\begin{gather*}
	(V-I, I)_{\rm L} = (3.036 , 21.551) \pm (0.743, 0.412)\ \ \ \ {\rm for\ small}\ \pi_{\rm E}, \\
	(V-I, I)_{\rm L}  = (3.756 , 22.318) \pm (0.193, 0.133)\ \ \ \ {\rm for\ large}\ \pi_{\rm E}.
\end{gather*}

It should be noted that these are estimates from the light-curve model, which includes a constraint on the total flux on the source and lens from the HST data. The lens color and magnitude from the light curve model are not significantly different from the  values measured for the lens in the HST follow-up data (see Table \ref{hst}). We plot these values, and of the two sources, on a color-magnitude diagram (CMD) in Figure \ref{fig:cmd}. The degeneracy in the microlensing parallax does not have a significant effect on the primary source's color and magnitude. However, the secondary source and the lens are impacted by the degenaracy, as seen on Figure \ref{fig:cmd}. 
	
The higher order effects that were included in the light-curve modeling of this event are discussed further in Section \ref{sec:highordereff}.

\subsection{Higher-order effects}
\label{sec:highordereff}

\subsubsection{Microlensing Parallax}
\label{sec:parallax}

The microlensing parallax, $\pi_{\rm E}$, is a 2D vector with an East and North component, $\pi_{\rm E,E}$ and $\pi_{\rm E,N}$, respectively \citep{Gould2000}. The orbital microlensing parallax effect occurs when the assumption that the lens moves at constant velocity with respect to the source (as seen by observers on Earth) breaks down. It is often detected for events with long time scales, $t_{\rm E} > 50$ days. Other types of parallax included terrestrial parallax and satellite parallax, although these cases are not relevant to this paper.

\subsubsection{Orbital Motion of Lens System}
\label{sec:orbital}

Orbital motion of the lens system is always present, however, their positions are often assumed to be fixed during the microlensing event. This is because a microlensing event's duration is often on the order of a few days to weeks, making is a small fraction of the lens systems' orbital period of several years. In our model we assume a face-on circular orbit with the parameters of orbital angular speed, $\omega$,  and the binary separation rate of $ds/dt$. If microlensing parallax effects are present on the light-curve, then it is physically realistic that the model include orbital motion effects too. However, due to rotational symmetry, $\pi_{\rm E,E}$ (which is the component perpendicular to the instantaneous direction of the Earth's acceleration) is often degenerate with $\omega$ \citep{Batista2011, Skowron2011}.

\subsubsection{Binary Source}
\label{sec:binary}

Since the original study by \cite{Furusawa2013} indicates the possibility of a second source, and our measured Keck K magnitude is brighter than that predicted, our model includes a binary lens and a binary source (2L2S). In the case of MB10328 we have binary source effects (i.e. magnification of the second source) and the xallarap effect.

The parameters due to the binary source effect are the time of closest approach  ($t_{0,2}$) and impact parameter ($u_{0,2}$) of the second source. Additionally, we have the $I$ and $V$ band flux ratios between the second source ($S_2$) and the primary source ($S_1$), fr2I and fr2V, respectively. The xallarap effect is when the source orbital motion due to the second source invokes the parameters $\rm dt_{\rm E21}$, which is implied by $t_{\rm E2} - t_{\rm E1}$; and $d\alpha$, implied by $\alpha_{\rm E2} - \alpha_{\rm E1}$.  In the light-curve modeling we also include the inverse of the source star orbital period, $1/T_{\rm Sbin}$.

\setlength{\tabcolsep}{6pt}
\begin{deluxetable*}{ l c c c c c}
	\tablecaption{Light-curve model Parameters using the microlensing parallax plus orbital motion close and wide solution, and the xallarap solution from \cite{Furusawa2013}; and light-curve model parameters from this work, separated into the `small' and `large' $\pi_{\rm E}$ parallax degeneracy. The new model includes microlensing parallax, orbital motion, xallarap, and magnification of a second source. Parameter descriptions can be found throughout Section \ref{lightcurvesection}. \label{tab-mparams2} }
	\tablewidth{0pt}
	\tablehead{
		&  \multicolumn{2}{c}{Parallax plus orbital motion } & & \multicolumn{2}{c}{This Work} \\
		\colhead{Parameter} 	 &   ($u_0 < 0$)  &($u_0 > 0$) &\colhead{Xallarap} &  small $\pi_{\rm E}$&large $\pi_{\rm E}$
	}  % end header.
	%\hline
	%\hline
	\startdata
	$t_{\rm E}$ (days) & $62.6\pm 0.6$ &$64.2 \pm 0.6$& $61.8 \pm 0.3$ & $61.24 \pm 1.37$  & 63.51 $\pm$ 0.84 \\   
	$t_0$ (${\rm HJD}^\prime)$&$5378.683 \pm 0.014$ & $5378.694 \pm 0.017$& $5378.706 \pm 0.013$ & 5378.737$\pm$ 0.521 & 5377.864 $\pm$ 0.194 \\
	$t_{*}$ (days) & 0.0582 $\pm$ 0.0063 & 0.0700 $\pm$ 0.0109 & 0.0834 $\pm$ 0.0007 & 0.0926$\pm$ 0.0104 & 0.0652 $\pm$ 0.0021\\
	$u_0$  &$- 0.0721 \pm 0.0008$  & $0.0716 \pm 0.0007$ & $- 0.0741 \pm 0.0004$& $- 0.0661 \pm 0.0019$ & $- 0.0507 \pm 0.0019$\\
	s  & $1.154 \pm 0.016$ & $1.180 \pm 0.028$ &$ 1.220 \pm 0.012$& 1.222 $\pm$ 0.006 & 1.228 $\pm$ 0.004 \\
		q/$10^{-4}$  & $2.60 \pm 0.53$ & $3.68 \pm 1.26$  & $5.16 \pm 0.06$& 6.584 $\pm$ 1.453 & 3.090 $\pm$ 0.213\\
	$\alpha$ & $- 0.2743 \pm 0.0087$ & 0.1965 $\pm$ 0.0151& $- 0.2024 \pm 0.0004$& $- 2.9937 \pm 0.0011$ & $- 3.0272 \pm 0.0047$ \\
	$\pi_{\rm E,N}$ &1.01 $\pm$ 0.06 &0.72 $\pm$ 0.05 & --& 0.138 $\pm$ 0.019& 0.400 $\pm$ 0.036 \\
	$\pi_{\rm E,E}$ & $- 0.51 \pm 0.04$ & $- 0.39 \pm 0.03$& -- & $- 0.007 \pm 0.011$& $- 0.247 \pm 0.056$\\
	$ds/dt\times 10^{3}\ (\rm days^{-1})$ & $2.51 \pm 0.63$ & $1.41 \pm 1.16 $ & -- & $-0.0464 \pm 0.0864$ &$-0.0915\pm0.0288$\\
	$\omega\times 10^{3}\ \rm (rad/day)$  & $- 7.39 \pm 0.39$ & $- 1.39 \pm 0.60$& -- & $-0.25 \pm 1.05$& $2.60 \pm 0.30$ \\ 
	$t_{0,2}$ (HJD') & -- & -- & -- & 5339.858$\pm$ 12.292 & 5328.335 $\pm$ 9.694\\
	$u_{0,2}$ & --&-- & --& $- 0.210 \pm  0.989$ & 1.201 $\pm$ 0.099\\
	f2rI & -- & -- & -- & 0.339 $\pm$ 0.073 & 0.321$\pm$ 0.055 \\ 
	f2rV & -- & -- & -- & 0.399$\pm$ 0.118 & 0.434 $\pm$ 0.078\\
	$\rm dt_{\rm E21}$ (days) & -- & -- & -- &  8.460$\pm$ 7.907 &  0.065 $\pm$ 2.273\\
	$\rm d\alpha$ (mas) & -- & -- & -- & $- 0.0115 \pm 0.0584 $& $- 0.0126 \pm 0.0368$\\
	$1/T_{\rm Sbin}$ (day$\rm s^{-1}$) & -- & -- & -- & 0.000140 $\pm$ 0.00026 & -0.000013 $\pm$ 0.000074 \\
	$t_{\rm fix}$ (days) & - & -& -& 5406.0 &  5406.0\\
	$D_{\rm S}$ (kpc) & -- & -- &  & 7.680 $\pm$ 0.575 & 7.786 $\pm$ 0.516\\
	$\chi^2$ & 5657.75 & 5660.31 &5652.59 & 4654.90 & 4654.28\\
	$dof$ & 5660&5660 &5658 & 4682 &4682\\
	\enddata
\end{deluxetable*}	

	\begin{figure*}[h!]
	\centering
	\includegraphics[width=12cm]{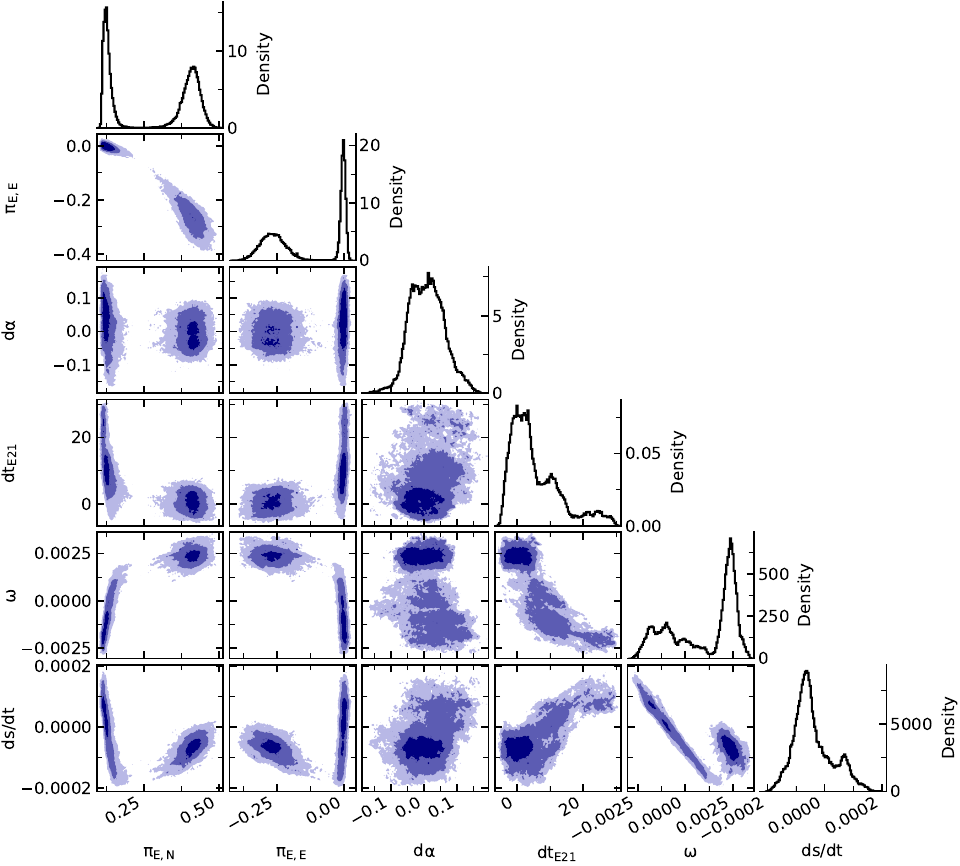}
	\caption{Corner plot from the MCMC probability distribution showing the parameters due to the microlenisng parallax, orbital motion, xallarap and magnification of the secondary source.}
	\label{fig:328corner}
\end{figure*}

	\begin{figure*}[h!]
	\centering
	\includegraphics[width=9cm]{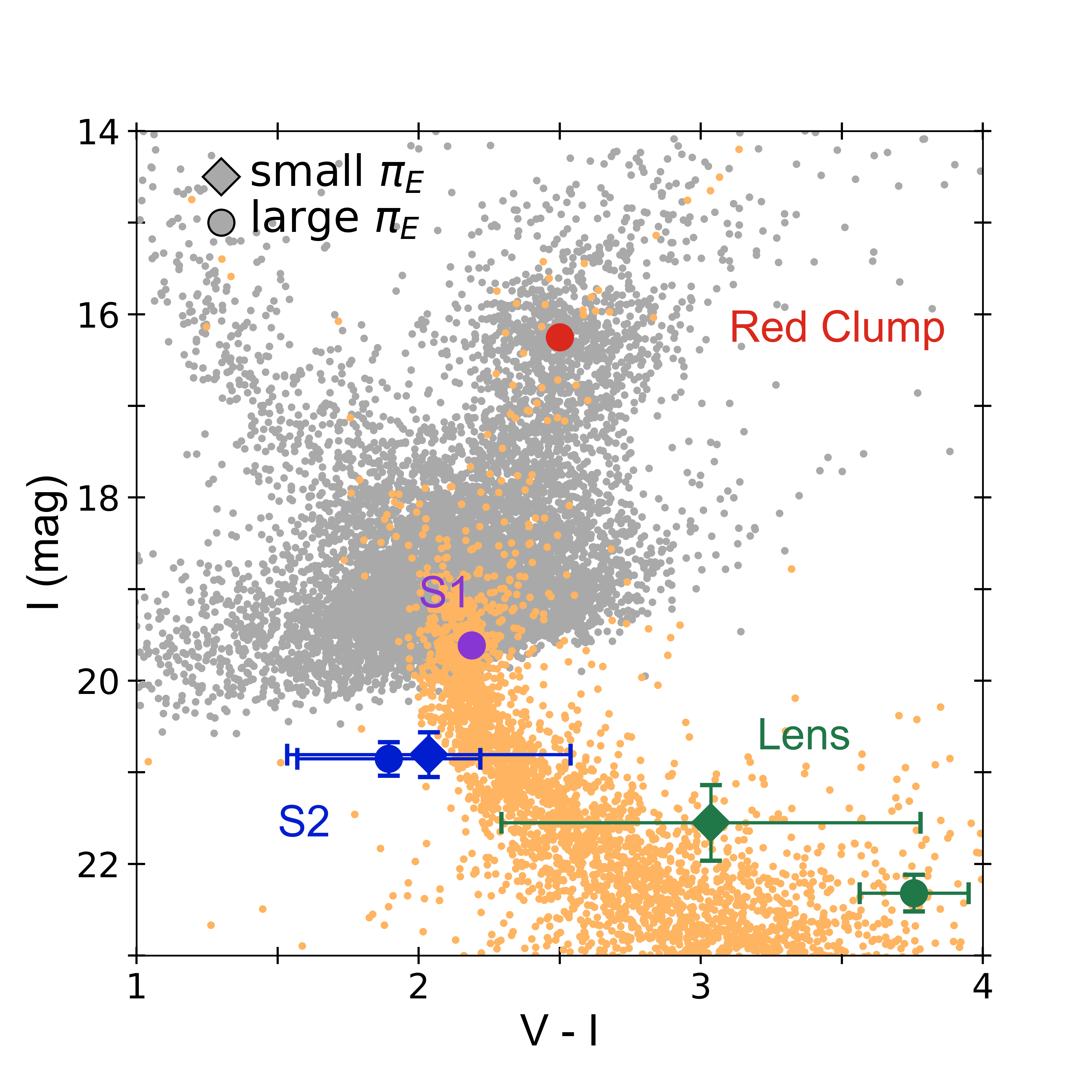}
	\caption{CMD of OGLE III stars within 120 arseconds of the MB10328 microlensing event, in grey. In orange is the color-magnitude diagram of the HST field from the 2018 epoch transformed to the same extinction and Galactic bar distance. The red clump giant centroid is indicated with a red dot. The purple marker is the primary source (S1), and the blue markers show the secondary source (S2) color and magntitude. The green markers represent the lens. The diamond markers indicate the values for the small $\pi_{\rm E}$ distribution, while the circles indicate values for the large $\pi_{\rm E}$. For S1, the difference and the error bars are too small to be seen on the figure. }
	\label{fig:cmd}
\end{figure*}

\section{Planetary System Parameters}
\label{sec:planet system}

Once the best-fit model parameters have been found, we can sum over the MCMC results presented in Table \ref{tab-mparams2}, and determine the posterior distribution of the physical properties of the lens system which are presented in Table \ref{tab:params}. We use light-curve parameters, in combination with the mass-distance and mass-luminosity relations in Equations \ref{eq:thetaE}, \ref{eq:pirel}, and \ref{eq:mL}. The constraints from the high-resolution follow-up data have already been applied in the light-curve modeling, and are therefore not applied again while summing over the MCMC results. Additionally, the source distance, $D_{\rm S}$, prior is also applied during the light-curve modeling, and therefore, not applied again. However, a Galactic model prior is applied on the lens distance for the $D_{\rm S}$ value. 
	
%% 
% the high resolution constraints, of lens-source relative proper motion mu, host star magnitude, and the combined lens + source magnitudes and color, are only applied in the light-curve modeling step, and do not need to be applied again. The source distance prior was applied to the light curve model - it is also not applied again. A galactic model prior, however, is used on the lens distance, for the Ds value in the light curve model (which is constrained by the measured mu_rel,H value.)
	
A common assumption used in the Bayesian analysis of microlensing events is that all stars are equally likely to host a planet with a fixed mass ratio, $q$. However, more recent studies (\cite{Koshimoto2021a, Bennett2024}) show that planet hosting probability of a fixed mass ratio increases linearly with the host mass. Therefore, this prior is included when summing over the MCMC results. Since we have constraints from high resolution imaging, this prior has a small effect on the results.  

Two mass-distance relations can be used which include the angular Einstein ring radius, $\theta_{\rm E}$, or the microlensing parallax $\pi_{\rm E}$ \citep{Beaulieu2008,Gaudi2012}. These are: 

\begin{equation}
	\label{eq:thetaE}
	M_{\rm L} = \frac{\theta_{\rm E}^2}{\kappa \pi_{\rm rel}},
\end{equation}

and

\begin{equation}
	\label{eq:pirel}
	M_{\rm L} = \frac{\pi_{\rm rel}}{\kappa \pi_{\rm E}^{2}},
\end{equation}

where $\pi_{\rm rel} = \rm AU(\frac{1}{D_{\rm L}} - \frac{1}{D_{\rm S}}),$ and $\kappa = 4G/(c^{2} AU) = 8.144\ \rm mas\ M_\odot^{-1}$. Therefore, with Equations \ref{eq:thetaE} and \ref{eq:pirel} an expression can be found for the lens mass $M_{\rm L}$, with no dependance on the lens or source distance, $D_{\rm L}$ or $D_{\rm S}$, respectively. The lens distance, however, does depend on the source distance. 

For the event MB10328, a direct measurement of the lens-source proper motion has been made, therefore $\theta_E$ can be constrained with $\theta_E = \mu_{\rm rel, G} t_E$. However, $\mu_{\rm rel, G}$ is in the instataneous geocentric inertial reference frame which moves with the earth at peak magnification. Since the proper motion derived from the high angular resolution follow up images are in the heliocentric frame, it needs to be converted to the geocentric one. This can be done using the equation: 
\begin{equation}
	\label{muG}
	{\boldmath \mu}_{\rm rel, G} = \mu_{\rm rel, Hel} - \frac{v_\oplus \pi_{\rm rel}}{\rm AU}
\end{equation}

Where $v_\oplus$ is the projected velocity of the earth relative to the sun at the time of the events peak magnification. For MB10328, this was $v_{\oplus E,N} = (28.878, -0.689)$ km/sec $= (6.088, -0.145)$ AU/yr at HJD' = 5406. Since $\pi_{\rm rel}$ is the relative parallax, we can rewrite equation \ref{muG} as: 
\begin{equation}
	{\boldmath \mu}_{\rm rel, G} = \mu_{\rm rel, Hel} - (6.088, -0.145) \times (1/D_{\rm L} - 1/D_{\rm S}),
\end{equation}

where $\mu_{\rm rel, Hel}$ and $\mu_{\rm rel, G}$ are in units of mas/yr and $D_{\rm L}$ and $D_{\rm S}$ are in units of kpc. This relation is used in our Bayesian analysis of the light-curve (Section \ref{lightcurvesection}). Therefore, from our Keck data we determine the geocentric relative proper motion to be $\mu_{\rm rel, G} = 4.208 \pm 0.061\ \rm mas\ yr^{-1}$ (for small $\pi_{\rm E}$), and $\mu_{\rm rel, G} = 5.119 \pm 0.162\ \rm mas\ yr^{-1}$ (for large $\pi_{\rm E}$).

%This translates to $\theta_{\rm E} = 0.701 \pm 0.020$ mas (for $\pi_{\rm E} < cut$), and $\theta_{\rm E} = 0.899 \pm 0.032$ mas (for $\pi_{\rm E} > cut$). 

The angular Einstein radius, $\theta_{\rm E}$, can be calculated by two methods. First by using light-curve model parameters, $\theta_{\rm E}$ can be calculated with $\theta_{\rm E} = \theta_{*}/\rho$, where $\rho = t_*/t_{\rm E}$ and $\theta_{*}$ are found from the color-surface brightness relation in Equation \ref{eq:surface brightness}; and secondly, from the  measured lens-source proper motion in the follow-up data, once it has been converted to the geocentric reference frame.The final $\theta_{\rm E}$ measurement, presented in Table \ref{tab:params}, comes from the weighted sum of these two methods, using the lens-source proper motion measured from Keck; the lens-source relative proper motion light-curve model parameters $t_{*}$, $t_{\rm E}$; and the source $V$ and $I$ magnitude using a color-surface brightness relation \citep{Boyajian2013}.

%\textbf{We can compare these $\theta_{\rm E}$ values to those calculated using parameters from the light-curve, using $\theta_{\rm E} = \theta_{*}/\rho$, where $\rho = t_*/t_{\rm E}$. For $\pi_{\rm E} < cut$ we calculate $\theta_{\rm E} = 0.558 \pm 0.064$ mas, and for $\pi_{\rm E} > cut$, $\theta_{\rm E} = 0.878 \pm 0.049$ mas.  The values for $\pi_{\rm E} > cut$ have $< 1\sigma$ agreement, however, for $\pi_{\rm E} < cut$ values the difference is $> 2\sigma$. }
 
A mass-luminosity relation can be used if the lens magnitude is measured. This can be done with high angular resolution follow-up observations where the source and lens are either fully resolved \citep{Bennett2015, Batista2015, Vandorou2020} or partially resolved, \citep{Bhattacharya2018, Bennett2020, Terry2022}.
\begin{equation}
	\label{eq:mL}
	m_{\rm L} = 10+5\log_{10}(D_{\rm L}/1\ \rm kpc) + A_{K,L}+ M_{\mathcal{K}}( M_{L}),
\end{equation}

where $M_{\mathcal{K}}(M_{\rm L})$ is a $K$ band luminosity relation. The interstellar K band extinction along the lens line of sight is given by $A_{\rm K,L}$. For this event the extinction we use is $A_{\rm K,L} = 0.185$ based on the \cite{Surot2020} E(J-K) map and \cite{Nishiyama2006} IR extinction law.  The empirical mass-luminosity relations are from \cite{Henry1993} ($M_{\rm L} \geqslant 0.66\ M_{\odot}$), \cite{Delfosse2000} ($0.54\ M_{\odot} \geqslant M_{\rm L} \geqslant 0.12\ M_{\odot}$) and \cite{Henry1999} ($0.10\ M_{\odot} \geqslant M_{\rm L} \geqslant 0.07\ M_{\odot}$). They are also used in \cite{Bennett2015, Bennett_2016, Bennett2018}. These three mass-luminosity relations are merged with a smooth interpolation between them.

\begin{figure*}[t!]
	\centering
	\includegraphics[width=18.5cm]{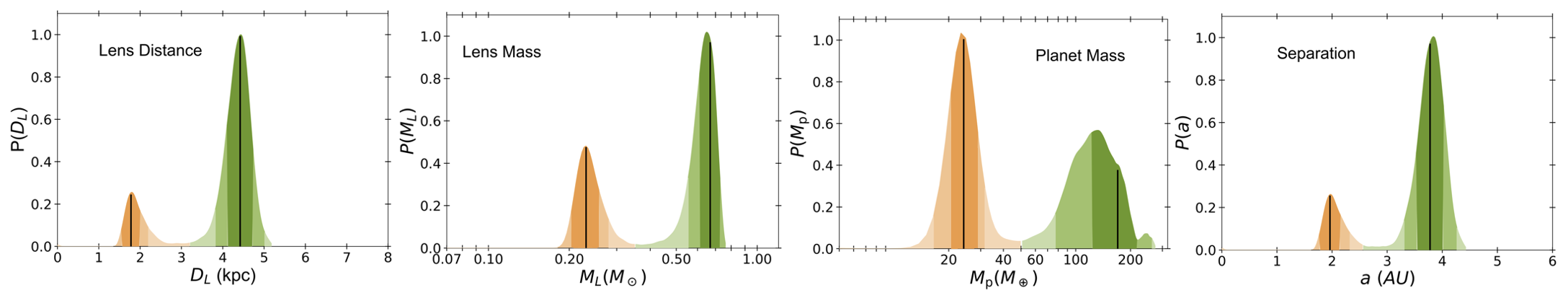}
	\caption{Bayesian probability distributions of the planetary system, including host mass and distance, separation and planetary companion mass. Orange represents $ \pi_{\rm E,N} > \pi_{\rm E,E} - 0.33$ values, and green respresents $\pi_{\rm E,N} < \pi_{\rm E,E} - 0.33$ values. The central 68.3\%  of the distributions are shaded in darker hues (of either orange or green) and the remaining central 95.4\% of the distributions are shaded in lighter hues. The vertical black line marks the median of the probability distribution of the respective parameters for the large and small $\pi_{\rm E}$ values.}
	\label{fig:328histogram}
\end{figure*}

By summing over the image-constrained light-curve model parameters and using the mass-distance and mass-luminosity relations, we determine the posterior distribution of the physical properties of the lens system, which are presented in Table \ref{tab:params}.  However, as we expected the distribution of the lens system's physical parameters is not a single Gaussian, and when looking at the probability distributions in Figure \ref{fig:328histogram} we see a clear degeneracy, which we investigate in more detail in the following section \ref{deg}.

\setlength{\tabcolsep}{10pt}
\begin{deluxetable}{ l c c c c c}
	\tablecaption{Planetary system parameters from high angular resolution follow up constraints compared to those found by \citet{Furusawa2013}.We show the lens system parameters for the large and small $\pi_{\rm E}$ values.}.  \label{tab:params} 
	\tablewidth{0pt}
	\tablehead{
		&  \multicolumn{2}{c}{Parallax plus orbital motion } & & \multicolumn{2}{c}{This Work}\\
		Parameter	 &   ($u_0 < 0$)  &($u_0 > 0$) & \colhead{Xallarap} & \colhead{small $\pi_{\rm E}$} & \colhead{large $\pi_{\rm E}$}
	}  % end header.
	%\hline
	%\hline
	\startdata
	$\theta_{\rm E}$ (mas) & 0.98 $\pm$ 0.12 & 0.83 $\pm$ 0.14 & 0.68 $\pm$ 0.04 & 0.713 $\pm$ 0.021 & 0.881 $\pm$ 0.347 \\
	$\mu_{\rm rel, G}$ ($\rm mas\ yr^{-1}$)& 5.71 $\pm$ 0.70 & 4.72 $\pm$ 0.79& 4.03 $\pm$ 0.26 &4.246$\pm$ 0.068 & 5.067 $\pm$ 0.191 \\
	Lens Mass ($M_{\odot}$) & 0.11 $\pm$ 0.01  & 0.12 $\pm$ 0.02 &  $0.64^{+0.22}_{-0.34}$&  0.633 $\pm$ 0.062 & 0.244 $\pm$ 0.030 \\   
	Lens Distance (kpc) & 0.81 $\pm$ 0.10 &1.24 $\pm$ 0.18 & $4.6^{+1.1}_{-1.8}$& 4.350 $\pm$ 0.320 & 1.928 $\pm$ 0.269\\
	Planet Mass ($M_{\rm Earth}$) & 9.2 $\pm$ 2.2  & 15.2 $\pm$ 5.9 & $109^{+38}_{-58}$ &141.13 $\pm$ 42.42 & 24.75 $\pm$ 3.39\\
	$\alpha_\perp$(AU)  & 0.92 $\pm$ 0.16 & 1.21 $\pm$ 0.27 & $3.8^{+0.9}_{-1.5}$&3.787 $\pm$ 0.242 & 2.075 $\pm$ 0.210\\
	\enddata
\end{deluxetable}	

\subsection{Investigating the degenerecy}
\label{deg}

In Figure \ref{fig:328histogram} we see two possible solutions with either an M-dwarf host of $\sim 0.2\ M_{\odot}$ at $\sim 2$ kpc or a higher mass one with $\sim 0.5\ M_{\odot}$ at $\sim 4$ kpc. This degeneracy is characteristic for M-dwarfs in the disk where different host masses can be challenging to distinguish since an early M-dwarf further away can appear to be a late M-dwarf that is closer. The planetary companion's mass is also affected. If the host is a late M-dwarf at a closer distance it is likely to host a sub-Neptune mass planet, if the host is a early M-dwarf further away, it likely hosts a Gas Giant with a mass $\sim2\ M_{\rm Saturn}$. The projected separation for each solution can be calculated with $a_{\perp} = sD_{\rm L}\theta_{\rm E}$. For both solutions, the planetary companion is likely to be beyond the snow line.

This event also exhibits a xallarap signal and magnification from a companion to the source, whch further complicates the degeneracy in the mass-distance relationship. Even with high resolution observations where we directly measure the lens flux and relative proper motion, we still have this degeneracy using NIR data. The $V$ band magnitude of the lens should be able to break this degeneracy, however, the lens magnitude and position was not well constrained in the 2018 HST data. This is because the lens is very faint in that passband, and the issue is made worse by the small  separation between the source and lens only being $\sim35$ mas.

Due to the interplay between all the higher order effects of microlensing parallax signals, xallarap, magnification of the second source and planetary orbital motion ambiguity is created. This can be seen on the corner plot in Figure \ref{fig:328corner} of the higher order effects.  For the lens system's physical parameters this degeneracy can also be seen in Figure \ref{fig:328histogram}, where orange represents the ``larger" values ( $ \pi_{\rm E,N} > \pi_{\rm E,E} - 0.33$), and green represents the ``smaller" values  ($ \pi_{\rm E,N} < \pi_{\rm E,E} - 0.33$). 

We notice that the larger $\pi_{\rm E}$ values (in orange) produce a lower mass host star of $\sim 0.2\ M_\odot$ at a closer distance of $\sim 2$ kpc, while the smaller $\pi_{\rm E}$ values produce a larger mass host with $\sim 0.7\ M_\odot$ further away, at a distance of $\sim 4.5$ kpc. This also affects the planet mass and separation as seen in Figure \ref{fig:328histogram}. We also present the physical parameters of the system in Table \ref{tab:params} for the ``large" and ``small" $\pi_{\rm E}$ values.

\begin{figure}[t!]
	\centering
	\includegraphics[width=18cm]{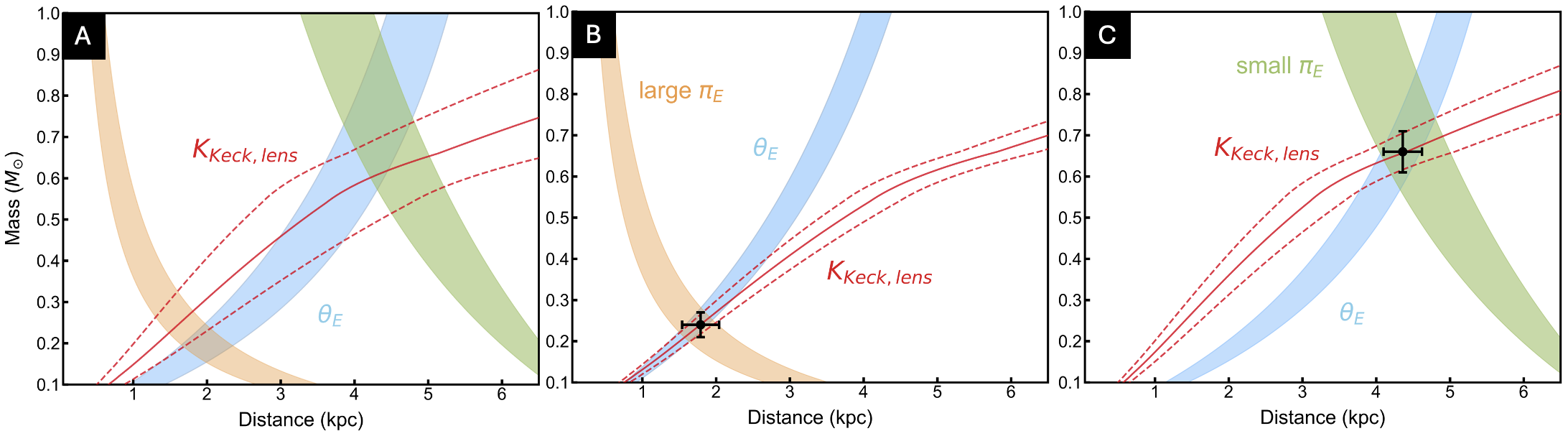}
	\caption{Mass-distance relations for MB10328, obtained using the Einstein ring radius, $\theta_E$ in blue and the Keck K-band flux constraint from mass-luminosity relations in red. Panel A shows the full MCMC probability distribution with the ``large" $\pi_{\rm E}$ constraint in orange and the ``small" $\pi_{\rm E}$ constraint in green. Panel B and C show the constraints for the ``large" and ``small" $\pi_{\rm E}$ values, along with the corresponding $\theta_E$ and K-band flux constraints. }
	\label{fig:328massdistance}
\end{figure}

This is also shown in Figure \ref{fig:328massdistance} which shows the mass-distance and mass-luminosity constraints from the high resolution follow-up data and described in Equations \ref{eq:thetaE}, \ref{eq:pirel} and \ref{eq:mL}. In this type of Figure, the intersection of different constraints results in the mass and distance of the lens star.  In Panel A of this Figure, however, we show that without the addition of a microlensing parallax measurement, there is a continuous degeneracy where the mass-distance constraint due to $\theta_E$ and that of due to the lens $K$ magnitude, do not intersect in any one place.  The addition of a parallax measurement should break this degeneracy, however, as we saw in Figure \ref{fig:328corner} there are two possible solutions for the microlensing parallax - which leads to two possible solutions for the lens mass and distance. We plot these separately in Panels B and C of Figure \ref{fig:328massdistance}

We also investigate the expected $V$ and $I$ magnitude of the lens associated with the large and small $\pi_{\rm E}$ values from the MCMC probability distribution. We find the following $2\sigma$ ranges ($25.68 < V_{\rm L} < 26.28$), ($21.96 < I_{\rm I} < 22.50$) and ($23.50 < V_{\rm L} < 25.59$), ($20.81 < I_{\rm I} < 22.18$) for the large and small $\pi_{\rm E}$ values, respectively. This means a V-band gap of 0.084 and an I-band gap of -0.221, which means this degeneracy can be broken with additional data in shorter-wavelengths, with telescopes such as HST or JWST.
 
\begin{figure}[t!]
	\centering
	\includegraphics[width=8cm]{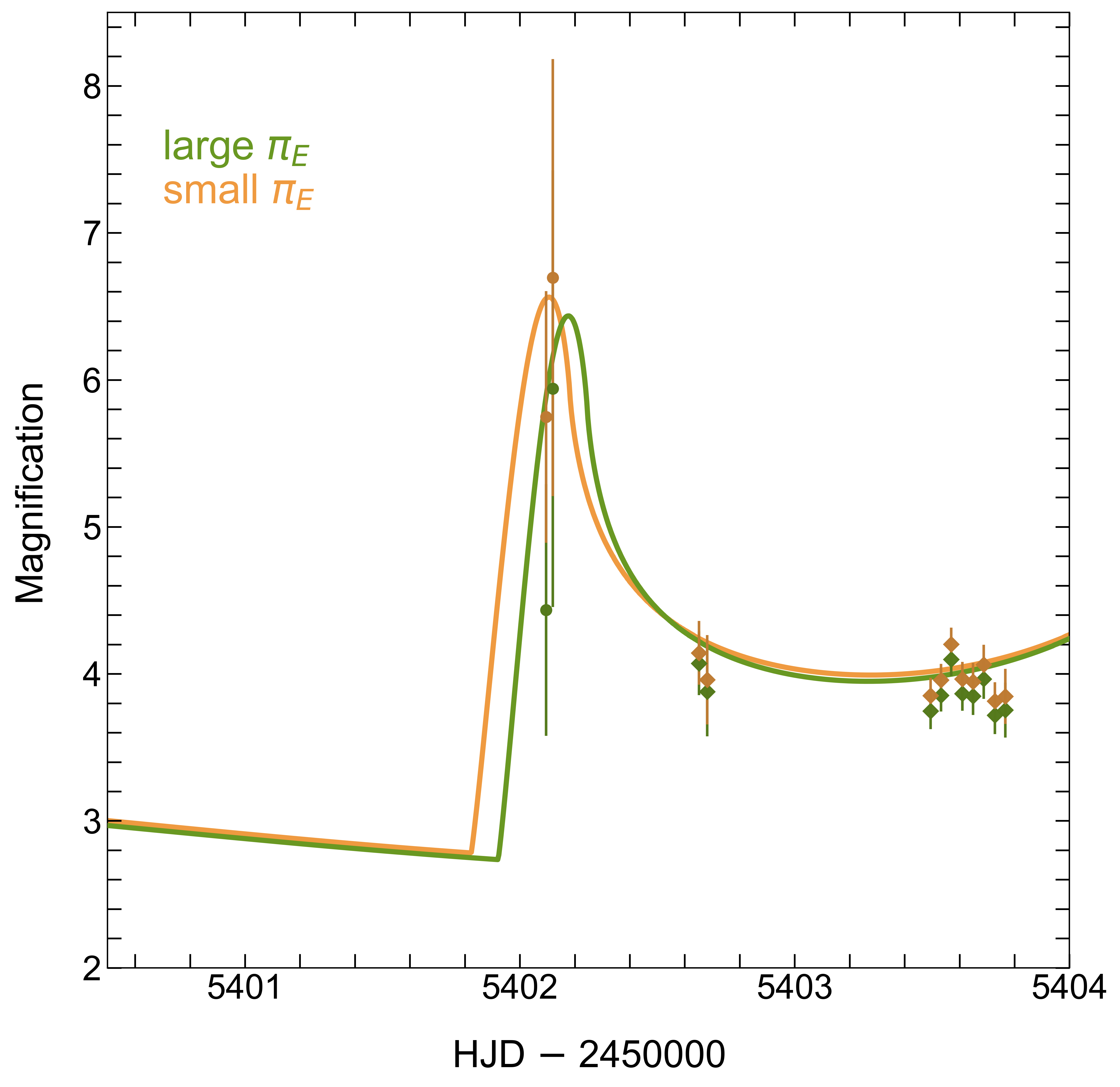}
	\caption{A zoom of the entry of the planetary anomaly for the two best-fit models, small $\pi_{\rm E}$ in orange and large $\pi_{\rm E}$ in green, and the corresponding MOA and OGLE data depicted by a circle and diamond, respectively. }
	\label{fig:LvsS}
\end{figure}

Another contributing factor to this degeneracy is the lack of coverage at the beginning of the planetary deviation between $\sim$5400 - 5406 days, which can be seen in the middle panel of Figure \ref{fig:lightcurve}.  In Figure \ref{fig:LvsS} we see that there is a small difference between the two models at the entrance of the planetary anomaly. The light-curve parameter affected the most by this is the host-planet mass ratio $q$, which is constrained by the shape and strength of the planetary anomaly.

	\section{Discussion and Conclusion }
	\label{section6}

	By using high angular resolution follow-up observations with Keck and HST we have been able to identify the MB10328 host star, measuring its $K$- band magnitude and relative proper motion to the source. We have also placed constraints on the total $I$ and $V$-band magnitude of the system. We determine that the host is an M-dwarf star in the disk either with a mass and distance of $\sim0.2\ M_{\odot}$ at 2-3 kpc, or $\sim0.5\ M_{\odot}$  at 4-5 kpc. This degeneracy was also seen in the detection paper by \cite{Furusawa2013} between the parallax plus orbital motion model and the xallarap model. Our high angular resolution follow-up results agreed with neither of these models. From the Keck data we detect a lens at the expected separation from the source, but there was additional stellar flux at the position of the source. Since a model with xallarap effects had already been explored in \cite{Furusawa2013}, it was clear that the source had a companion.
	
	Therefore, in this paper we remodel the light-curve using all these higher order effects; microlensing parallax, orbital motion, xallarap, and magnification of the second source. In the initial study by \cite{Furusawa2013} they explored higher order effects on the light-curve, but did not present a model that combined all of them. Instead, they investigated an orbital motion plus parallax model, and a xallarap model separately, finding these two to be degenerate. We also include constraints from the high resolution follow-up data from Keck and HST. From Keck we use the measured magnitude and direction of the relative source-lens proper motion as a constraint, and from HST we use the total (source+lens) measured $V$ and $I$ magnitudes as constraints. This new model was able to both match the positions and magnitudes of the lens and source stars, as seen in the Keck and HST data. The higher order constraints in this case play an important role when modeling the light-curve. Modeling of planetary systems is in general simple if constraints are applied, and typically acts to improve the efficiency of the MCMC by excluding models that are inconsisted with the high resolution follow-up observations. However, in cases like MB10328 where parallax, orbital motion and xallarap are applied, without constraints from high resolution observations it is possible that the modeling code would be unable to find the correct solution.  The likelihood of a model to fail is increased when the higher order effect parameters are not constrained to physically reasonable values with a prior distribution.
	
		MB10328 is also part of the \cite{Suzuki2016} statistical sample, and these results will contribute to the determination of exoplanet host star masses. The \cite{Suzuki2016} sample will help us understand the dependence of planet occurance rate on the mass and distance of the host star. The analysis of other events in this sample have also highlighted the importance of including higher order effects on the light-curve. Our high angular resolution follow-up data (from Keck and HST) indicate that the modeling of these events had to be redone in order to determine the physical parameters of the system e.g. MOA-2007-BLG-192 \citep{Bennett2008, Terry2024} and OGLE-2012-BLG-0563 (\citet{Fukui2015, Bennett2024_0563}, Bhattacharya et al., \textit{submitted to AJ}).
	
	The inclusion of higher order effects like microlensing parallax in light-curve modeling is of particular importance. Every microlensing event, observed in a heliocentric reference frame will have microlensing parallax present, even if the effect is not measurable on the light-curve. Ignoring these effects could lead to errors in other light-curve parameters. This poses challenges in determining the physical parameters of microlensing systems, which complicates conducting statistical analyses of exoplanetary system properties from microlensing surveys, such as the future Roman Galactic Exoplanet Survey (RGES) \citep{Bennett2018, Penny2019, Johnson2020}. In this work we show the importance of including higher order effects on the light-curve and especially the use of image constrained modeling with high resolution follow-up data. This method will play an important role for the RGES to measure the masses and distances of all the exoplanetary systems it detects.
	
		In the case of MB10328 the solution is still ambiguous even with a clear lens detection in the Keck $K$ band data. Our solution indicates that the system could either be a nearby late M-dwarf with a mini-Neptune companion, or a more distant early M-dwarf with a Gas Giant companion. This degeneracy is not unusual for M-dwarfs in the disk, and can be broken with observations in shorter wavelengths. HST data for this event taken in 2018, however, could not robustly detect the lens due to its faintness relative to the source and because of the small separation. Additional data taken with either HST or JWST now that the relative source-lens separation is larger,  could break this degenracy and provide a clear solution to this event once and for all.

	\acknowledgments

\footnotesize{This work is supported by NASA through grant NASA-80NSSC18K0274 and by the University of Tasmania through the UTAS Foundation and the endowed Warren Chair in Astronomy and the ANR COLD-WORLDS (ANR-18-CE31-0002. This research was also supported in part by the Australian Government through the Australian Research Council Discovery Program (project number 200101909) grant awarded to AC and JPB. The Keck Telescope observations and analysis were supported by a NASA Keck PI Data Award, administered by the NASA Exoplanet Science Institute. Data presented herein were obtained at the W. M. Keck Observatory from telescope time allocated to NASA through the agency’s scientific partnership with the California Institute of Technology and the University of California. The Keck I and Keck II real-time controller upgrades are funded by the National Science Foundation’s Mid-Scale Innovations Program award AST-1836016 and Major Research for Instrumentation Program award AST-1727071, respectively. The Observatory was made possible by the generous financial support of the W. M. Keck Foundation. The authors wish to recognize and acknowledge the very significant cultural role and reverence that the summit of Mauna Kea has always had within the indigenous Hawaiian community. We are most fortunate to have the opportunity to conduct observations from this mountain. DPB and AB were also supported by NASA through grant NASA-80NSSC18K0274.}

		\bibliographystyle{yahapj}
	\bibliography{lib}

\begin{thebibliography}{}
\providecommand\natexlab[1]{#1}
\providecommand\JournalTitle[1]{#1}

\bibitem[{{Batista} {et~al.}(2015){Batista}, {Beaulieu}, {Bennett}, {Gould},
  {Marquette}, {Fukui}, \& {Bhattacharya}}]{Batista2015}
{Batista}, V., {Beaulieu}, J.~P., {Bennett}, D.~P., {et~al.} 2015,
  \href{http://dx.doi.org/10.1088/0004-637X/808/2/170}{\JournalTitle{\apj},
  808, 170}

\bibitem[{{Batista} {et~al.}(2011){Batista}, {Gould}, {Dieters}, {Dong},
  {Bond}, {Beaulieu}, {Maoz}, {Monard}, {Christie}, {McCormick}, {Albrow},
  {Horne}, {Tsapras}, {Burgdorf}, {Calchi Novati}, {Skottfelt}, {Caldwell},
  {Koz{\l}owski}, {Kubas}, {Gaudi}, {Han}, {Bennett}, {An}, {MOA
  Collaboration}, {Abe}, {Botzler}, {Douchin}, {Freeman}, {Fukui}, {Furusawa},
  {Hearnshaw}, {Hosaka}, {Itow}, {Kamiya}, {Kilmartin}, {Korpela}, {Lin},
  {Ling}, {Makita}, {Masuda}, {Matsubara}, {Miyake}, {Muraki}, {Nagaya},
  {Nishimoto}, {Ohnishi}, {Okumura}, {Perrott}, {Rattenbury}, {Saito},
  {Sullivan}, {Sumi}, {Sweatman}, {Tristram}, {von Seggern}, {Yock}, {PLANET
  Collaboration}, {Brillant}, {Calitz}, {Cassan}, {Cole}, {Cook}, {Coutures},
  {Dominis Prester}, {Donatowicz}, {Greenhill}, {Hoffman}, {Jablonski}, {Kane},
  {Kains}, {Marquette}, {Martin}, {Martioli}, {Meintjes}, {Menzies},
  {Pedretti}, {Pollard}, {Sahu}, {Vinter}, {Wambsganss}, {Watson}, {Williams},
  {Zub}, {FUN Collaboration}, {Allen}, {Bolt}, {Bos}, {DePoy}, {Drummond},
  {Eastman}, {Gal-Yam}, {Gorbikov}, {Higgins}, {Janczak}, {Kaspi}, {Lee},
  {Mallia}, {Maury}, {Monard}, {Moorhouse}, {Morgan}, {Natusch}, {Ofek},
  {Park}, {Pogge}, {Polishook}, {Santallo}, {Shporer}, {Spector}, {Thornley},
  {Yee}, {MiNDSTEp Consortium}, {Bozza}, {Browne}, {Dominik}, {Dreizler},
  {Finet}, {Glitrup}, {Grundahl}, {Harps{\o}e}, {Hessman}, {Hinse},
  {Hundertmark}, {J{\o}rgensen}, {Liebig}, {Maier}, {Mancini}, {Mathiasen},
  {Rahvar}, {Ricci}, {Scarpetta}, {Southworth}, {Surdej}, {Zimmer}, {RoboNet
  Collaboration}, {Allan}, {Bramich}, {Snodgrass}, {Steele}, \&
  {Street}}]{Batista2011}
{Batista}, V., {Gould}, A., {Dieters}, S., {et~al.} 2011,
  \href{http://dx.doi.org/10.1051/0004-6361/201016111}{\JournalTitle{AAP}, 529,
  A102}

\bibitem[{{Batista} {et~al.}(2014){Batista}, {Beaulieu}, {Gould}, {Bennett},
  {Yee}, {Fukui}, {Gaudi}, {Sumi}, \& {Udalski}}]{Batista2014}
{Batista}, V., {Beaulieu}, J.-P., {Gould}, A., {et~al.} 2014,
  \JournalTitle{ApJ}, 780

\bibitem[{Beaulieu(2018)}]{Beaulieu2018}
Beaulieu, J.-P. 2018,
  \href{http://www.mdpi.com/2218-1997/4/4/61}{\JournalTitle{Universe}, 4}

\bibitem[{{Beaulieu} {et~al.}(2006){Beaulieu}, {Bennett}, {Fouqu{\'e}},
  {Williams}, {Dominik}, {J{\o}rgensen}, {Kubas}, {Cassan}, {Coutures},
  {Greenhill}, {Hill}, {Menzies}, {Sackett}, {Albrow}, {Brillant}, {Caldwell},
  {Calitz}, {Cook}, {Corrales}, {Desort}, {Dieters}, {Dominis}, {Donatowicz},
  {Hoffman}, {Kane}, {Marquette}, {Martin}, {Meintjes}, {Pollard}, {Sahu},
  {Vinter}, {Wambsganss}, {Woller}, {Horne}, {Steele}, {Bramich}, {Burgdorf},
  {Snodgrass}, {Bode}, {Udalski}, {Szyma{\'n}ski}, {Kubiak},
  {Wi{\c{e}}ckowski}, {Pietrzy{\'n}ski}, {Soszy{\'n}ski}, {Szewczyk},
  {Wyrzykowski}, {Paczy{\'n}ski}, {Abe}, {Bond}, {Britton}, {Gilmore},
  {Hearnshaw}, {Itow}, {Kamiya}, {Kilmartin}, {Korpela}, {Masuda}, {Matsubara},
  {Motomura}, {Muraki}, {Nakamura}, {Okada}, {Ohnishi}, {Rattenbury}, {Sako},
  {Sato}, {Sasaki}, {Sekiguchi}, {Sullivan}, {Tristram}, {Yock}, \&
  {Yoshioka}}]{Beaulieu2006}
{Beaulieu}, J.~P., {Bennett}, D.~P., {Fouqu{\'e}}, P., {et~al.} 2006,
  \href{http://dx.doi.org/10.1038/nature04441}{\JournalTitle{\nat}, 439, 437}

\bibitem[{{Beaulieu} {et~al.}(2008){Beaulieu}, {Kerins}, {Mao}, {Bennett},
  {Cassan}, {Dieters}, {Gaudi}, {Gould}, {Batista}, {Bender}, {Brillant},
  {Cook}, {Coutures}, {Dominis-Prester}, {Donatowicz}, {Fouqu{\'e}}, {Grebel},
  {Greenhill}, {Heyrovsky}, {Horne}, {Kubas}, {Marquette}, {Menzies},
  {Rattenbury}, {Ribas}, {Sahu}, {Tsapras}, {Udalski}, \&
  {Vinter}}]{Beaulieu2008}
{Beaulieu}, J.~P., {Kerins}, E., {Mao}, S., {et~al.} 2008, \JournalTitle{arXiv
  e-prints}, arXiv:0808.0005

\bibitem[{{Beaulieu} {et~al.}(2016){Beaulieu}, {Bennett}, {Batista}, {Fukui},
  {Marquette}, {Brillant}, {Cole}, {Rogers}, {Sumi}, {Abe}, {Bhattacharya},
  {Koshimoto}, {Suzuki}, {Tristram}, {Han}, {Gould}, {Pogge}, \&
  {Yee}}]{Beaulieu2016}
{Beaulieu}, J.~P., {Bennett}, D.~P., {Batista}, V., {et~al.} 2016,
  \href{http://dx.doi.org/10.3847/0004-637X/824/2/83}{\JournalTitle{\apj}, 824,
  83}

\bibitem[{{Bennett}(2019)}]{BennettWFIRST}
{Bennett}, D. 2019, {Development of the WFIRST Exoplanet Mass Measurement
  Method}, Keck Observatory Archive N021

\bibitem[{{Bennett} {et~al.}(2006){Bennett}, {Anderson}, {Bond}, {Udalski}, \&
  {Gould}}]{Bennett2006}
{Bennett}, D.~P., {Anderson}, J., {Bond}, I.~A., {Udalski}, A., \& {Gould}, A.
  2006, \href{http://dx.doi.org/10.1086/507585}{\JournalTitle{\apjl}, 647,
  L171}

\bibitem[{{Bennett} {et~al.}(2007){Bennett}, {Anderson}, \&
  {Gaudi}}]{Bennett2007}
{Bennett}, D.~P., {Anderson}, J., \& {Gaudi}, B.~S. 2007,
  \href{http://dx.doi.org/10.1086/513013}{\JournalTitle{\apj}, 660, 781}

\bibitem[{Bennett \& Rhie(1996)}]{Bennett1996}
Bennett, D.~P., \& Rhie, S.~H. 1996,
  \href{http://dx.doi.org/10.1086/178096}{\JournalTitle{ApJ}, 472, 660}

\bibitem[{{Bennett} {et~al.}(2008){Bennett}, {Bond}, {Udalski}, {Sumi}, {Abe},
  {Fukui}, {Furusawa}, {Hearnshaw}, {Holderness}, {Itow}, {Kamiya}, {Korpela},
  {Kilmartin}, {Lin}, {Ling}, {Masuda}, {Matsubara}, {Miyake}, {Muraki},
  {Nagaya}, {Okumura}, {Ohnishi}, {Perrott}, {Rattenbury}, {Sako}, {Saito},
  {Sato}, {Skuljan}, {Sullivan}, {Sweatman}, {Tristram}, {Yock}, {Kubiak},
  {Szyma{\'n}ski}, {Pietrzy{\'n}ski}, {Soszy{\'n}ski}, {Szewczyk},
  {Wyrzykowski}, {Ulaczyk}, {Batista}, {Beaulieu}, {Brillant}, {Cassan},
  {Fouqu{\'e}}, {Kervella}, {Kubas}, \& {Marquette}}]{Bennett2008}
{Bennett}, D.~P., {Bond}, I.~A., {Udalski}, A., {et~al.} 2008,
  \href{http://dx.doi.org/10.1086/589940}{\JournalTitle{\apj}, 684, 663}

\bibitem[{{Bennett} {et~al.}(2010){Bennett}, {Rhie}, {Nikolaev}, {Gaudi},
  {Udalski}, {Gould}, {Christie}, {Maoz}, {Dong}, {McCormick}, {Szyma{\'n}ski},
  {Tristram}, {Macintosh}, {Cook}, {Kubiak}, {Pietrzy{\'n}ski},
  {Soszy{\'n}ski}, {Szewczyk}, {Ulaczyk}, {Wyrzykowski}, {OGLE Collaboration},
  {DePoy}, {Han}, {Kaspi}, {Lee}, {Mallia}, {Natusch}, {Park}, {Pogge},
  {Polishook}, {{\ensuremath{\mu}}FUN Collaboration}, {Abe}, {Bond}, {Botzler},
  {Fukui}, {Hearnshaw}, {Itow}, {Kamiya}, {Korpela}, {Kilmartin}, {Lin},
  {Ling}, {Masuda}, {Matsubara}, {Motomura}, {Muraki}, {Nakamura}, {Okumura},
  {Ohnishi}, {Perrott}, {Rattenbury}, {Sako}, {Saito}, {Sato}, {Skuljan},
  {Sullivan}, {Sumi}, {Sweatman}, {Yock}, {MOA Collaboration}, {Albrow},
  {Allan}, {Beaulieu}, {Bramich}, {Burgdorf}, {Coutures}, {Dominik}, {Dieters},
  {Fouqu{\'e}}, {Greenhill}, {Horne}, {Snodgrass}, {Steele}, {Tsapras},
  {PLANET}, {RoboNet Collaborations}, {Chaboyer}, {Crocker}, \&
  {Frank}}]{Bennett2010}
{Bennett}, D.~P., {Rhie}, S.~H., {Nikolaev}, S., {et~al.} 2010,
  \href{http://dx.doi.org/10.1088/0004-637X/713/2/837}{\JournalTitle{\apj},
  713, 837}

\bibitem[{{Bennett} {et~al.}(2015){Bennett}, {Bhattacharya}, {Anderson},
  {Bond}, {Anderson}, {Barry}, {Batista}, {Beaulieu}, {DePoy}, {Dong}, {Gaudi},
  {Gilbert}, {Gould}, {Pfeifle}, {Pogge}, {Suzuki}, {Terry}, \&
  {Udalski}}]{Bennett2015}
{Bennett}, D.~P., {Bhattacharya}, A., {Anderson}, J., {et~al.} 2015,
  \href{http://dx.doi.org/10.1088/0004-637X/808/2/169}{\JournalTitle{\apj},
  808, 169}

\bibitem[{Bennett {et~al.}(2016)Bennett, Rhie, Udalski, Gould, Tsapras, Kubas,
  Bond, Greenhill, Cassan, Rattenbury, Boyajian, Luhn, Penny, Anderson, Abe,
  Bhattacharya, Botzler, Donachie, Freeman, Fukui, Hirao, Itow, Koshimoto, Li,
  Ling, Masuda, Matsubara, Muraki, Nagakane, Ohnishi, Oyokawa, Perrott, Saito,
  Sharan, Sullivan, Sumi, Suzuki, Tristram, Yonehara, Yock, Collaboration),
  Szymański, Soszyński, Ulaczyk, Wyrzykowski, Collaboration), Allen, DePoy,
  Gal-Yam, Gaudi, Han, Monard, Ofek, Pogge, Collaboration), Street, Bramich,
  Dominik, Horne, Snodgrass, Steele, Collaboration), Albrow, Bachelet, Batista,
  Beaulieu, Brillant, Caldwell, Cole, Coutures, Dieters, Prester, Donatowicz,
  Fouqué, Hundertmark, Jørgensen, Kains, Kane, Marquette, Menzies, Pollard,
  Ranc, Sahu, Wambsganss, Williams, Zub, \& Collaboration)}]{Bennett_2016}
Bennett, D.~P., Rhie, S.~H., Udalski, A., {et~al.} 2016,
  \href{http://dx.doi.org/10.3847/0004-6256/152/5/125}{\JournalTitle{AJ}, 152,
  125}

\bibitem[{{Bennett} {et~al.}(2018){Bennett}, {Udalski}, {Bond}, {Suzuki},
  {Ryu}, {Abe}, {Barry}, {Bhattacharya}, {Donachie}, {Fukui}, {Hirao},
  {Kawasaki}, {Kondo}, {Koshimoto}, {Li}, {Matsubara}, {Miyazaki}, {Muraki},
  {Nagakane}, {Ohnishi}, {Ranc}, {Rattenbury}, {Suematsu}, {Sumi}, {Tristram},
  {Yonehara}, {MOA Collaboration}, {Szyma{\'n}ski}, {Soszy{\'n}ski},
  {Wyrzykowski}, {Ulaczyk}, {Poleski}, {Koz{\l}owski}, {Pietrukowicz},
  {Skowron}, {OGLE Collaboration}, {Shvartzvald}, {Maoz}, {Kaspi}, {Friedmann},
  {Wise Group}, {Batista}, {DePoy}, {Dong}, {Gaudi}, {Gould}, {Han}, {Pogge},
  {Tan}, {Yee}, \& {{\ensuremath{\mu}}FUN Collaboration}}]{Bennett2018}
{Bennett}, D.~P., {Udalski}, A., {Bond}, I.~A., {et~al.} 2018,
  \href{http://dx.doi.org/10.3847/1538-3881/aad59c}{\JournalTitle{AJ}, 156,
  113}

\bibitem[{{Bennett} {et~al.}(2020){Bennett}, {Bhattacharya}, {Beaulieu},
  {Blackman}, {Vandorou}, {Terry}, {Cole}, {Henderson}, {Koshimoto}, {Lu},
  {Baptiste Marquette}, {Ranc}, \& {Udalski}}]{Bennett2020}
{Bennett}, D.~P., {Bhattacharya}, A., {Beaulieu}, J.-P., {et~al.} 2020,
  \href{http://dx.doi.org/10.3847/1538-3881/ab6212}{\JournalTitle{\aj}, 159,
  68}

\bibitem[{{Bennett} {et~al.}(2024){Bennett}, {Bhattacharya}, {Beaulieu},
  {Koshimoto}, {Blackman}, {Bond}, {Ranc}, {Rektsini}, {Terry}, \&
  {Vandorou}}]{Bennett2024_0563}
---. 2024,
  \href{http://dx.doi.org/10.48550/arXiv.2412.03651}{\JournalTitle{arXiv
  e-prints}, arXiv:2412.03651}

\bibitem[{Bennett {et~al.}(2024)Bennett, Bhattacharya, Beaulieu, Koshimoto,
  Blackman, Bond, Ranc, Rektsini, Terry, Vandorou, Lu, Marquette, Olmschenk, \&
  Suzuki}]{Bennett2024}
Bennett, D.~P., Bhattacharya, A., Beaulieu, J.-P., {et~al.} 2024,
  \href{http://dx.doi.org/10.3847/1538-3881/ad4880}{\JournalTitle{ApJ}, 168,
  15}

\bibitem[{{Bertin}(2010)}]{Bertin2010}
{Bertin}, E. 2010, {SWarp: Resampling and Co-adding FITS Images Together},
  Astrophysics Source Code Library,
  \href{http://arxiv.org/abs/1010.068}{{\sffamily ascl:1010.068}}

\bibitem[{{Bertin} \& {Arnouts}(1996)}]{Bertin1996}
{Bertin}, E., \& {Arnouts}, S. 1996,
  \href{http://dx.doi.org/10.1051/aas:1996164}{\JournalTitle{\aaps}, 117, 393}

\bibitem[{{Bessell} \& {Brett}(1988)}]{Bessell1988}
{Bessell}, M.~S., \& {Brett}, J.~M. 1988,
  \href{http://dx.doi.org/10.1086/132281}{\JournalTitle{\pasp}, 100, 1134}

\bibitem[{{Bhattacharya} {et~al.}(2018){Bhattacharya}, {Beaulieu}, {Bennett},
  {Anderson}, {Koshimoto}, {Lu}, {Batista}, {Blackman}, {Bond}, {Fukui},
  {Henderson}, {Hirao}, {Marquette}, {Mroz}, {Ranc}, \&
  {Udalski}}]{Bhattacharya2018}
{Bhattacharya}, A., {Beaulieu}, J.~P., {Bennett}, D.~P., {et~al.} 2018,
  \href{http://dx.doi.org/10.3847/1538-3881/aaed46}{\JournalTitle{\aj}, 156,
  289}

\bibitem[{{Bhattacharya} {et~al.}(2021){Bhattacharya}, {Bennett}, {Beaulieu},
  {Bond}, {Koshimoto}, {Lu}, {Blackman}, {Vandorou}, {Terry}, {Batista},
  {Marquette}, {Cole}, {Fukui}, {Henderson}, \& {Ranc}}]{Bhattacharya2021}
{Bhattacharya}, A., {Bennett}, D.~P., {Beaulieu}, J.~P., {et~al.} 2021,
  \href{http://dx.doi.org/10.3847/1538-3881/abfec5}{\JournalTitle{\aj}, 162,
  60}

\bibitem[{{Blackman} {et~al.}(2021){Blackman}, {Beaulieu}, {Bennett},
  {Danielski}, {Alard}, {Cole}, {Vandorou}, {Ranc}, {Terry}, {Bhattacharya},
  {Bond}, {Bachelet}, {Veras}, {Koshimoto}, {Batista}, \&
  {Marquette}}]{Blackman2021Nature}
{Blackman}, J.~W., {Beaulieu}, J.~P., {Bennett}, D.~P., {et~al.} 2021,
  \href{http://dx.doi.org/10.1038/s41586-021-03869-6}{\JournalTitle{Nature},
  598, 272}

\bibitem[{{Borucki} {et~al.}(2011){Borucki}, {Koch}, {Basri}, {Batalha},
  {Brown}, {Bryson}, {Caldwell}, {Christensen-Dalsgaard}, {Cochran}, {DeVore},
  {Dunham}, {Gautier}, {Geary}, {Gilliland}, {Gould}, {Howell}, {Jenkins},
  {Latham}, {Lissauer}, {Marcy}, {Rowe}, {Sasselov}, {Boss}, {Charbonneau},
  {Ciardi}, {Doyle}, {Dupree}, {Ford}, {Fortney}, {Holman}, {Seager},
  {Steffen}, {Tarter}, {Welsh}, {Allen}, {Buchhave}, {Christiansen}, {Clarke},
  {Das}, {D{\'e}sert}, {Endl}, {Fabrycky}, {Fressin}, {Haas}, {Horch},
  {Howard}, {Isaacson}, {Kjeldsen}, {Kolodziejczak}, {Kulesa}, {Li}, {Lucas},
  {Machalek}, {McCarthy}, {MacQueen}, {Meibom}, {Miquel}, {Prsa}, {Quinn},
  {Quintana}, {Ragozzine}, {Sherry}, {Shporer}, {Tenenbaum}, {Torres},
  {Twicken}, {Van Cleve}, {Walkowicz}, {Witteborn}, \& {Still}}]{Borucki2011}
{Borucki}, W.~J., {Koch}, D.~G., {Basri}, G., {et~al.} 2011,
  \href{http://dx.doi.org/10.1088/0004-637X/736/1/19}{\JournalTitle{\apj}, 736,
  19}

\bibitem[{Boyajian {et~al.}(2014)Boyajian, van Belle, \& von
  Braun}]{Boyajian_2014}
Boyajian, T.~S., van Belle, G., \& von Braun, K. 2014,
  \href{http://dx.doi.org/10.1088/0004-6256/147/3/47}{\JournalTitle{ApJ}, 147,
  47}

\bibitem[{{Boyajian} {et~al.}(2013){Boyajian}, {von Braun}, {van Belle},
  {Farrington}, {Schaefer}, {Jones}, {White}, {McAlister}, {ten Brummelaar},
  {Ridgway}, {Gies}, {Sturmann}, {Sturmann}, {Turner}, {Goldfinger}, \&
  {Vargas}}]{Boyajian2013}
{Boyajian}, T.~S., {von Braun}, K., {van Belle}, G., {et~al.} 2013,
  \href{http://dx.doi.org/10.1088/0004-637X/771/1/40}{\JournalTitle{\apj}, 771,
  40}

\bibitem[{{Delfosse} {et~al.}(2000){Delfosse}, {Forveille}, {S{\'e}gransan},
  {Beuzit}, {Udry}, {Perrier}, \& {Mayor}}]{Delfosse2000}
{Delfosse}, X., {Forveille}, T., {S{\'e}gransan}, D., {et~al.} 2000,
  \JournalTitle{AAP}, 364, 217

\bibitem[{{Dominik} {et~al.}(2010){Dominik}, {J{\o}rgensen}, {Rattenbury},
  {Mathiasen}, {Hinse}, {Calchi Novati}, {Harps{\o}e}, {Bozza}, {Anguita},
  {Burgdorf}, {Horne}, {Hundertmark}, {Kerins}, {Kj{\ae}rgaard}, {Liebig},
  {Mancini}, {Masi}, {Rahvar}, {Ricci}, {Scarpetta}, {Snodgrass}, {Southworth},
  {Street}, {Surdej}, {Th{\"o}ne}, {Tsapras}, {Wambsganss}, \&
  {Zub}}]{Dominik2010}
{Dominik}, M., {J{\o}rgensen}, U.~G., {Rattenbury}, N.~J., {et~al.} 2010,
  \href{http://dx.doi.org/10.1002/asna.201011400}{\JournalTitle{Astronomische
  Nachrichten}, 331, 671}

\bibitem[{{Dong} {et~al.}(2009){Dong}, {Gould}, {Udalski}, {Anderson},
  {Christie}, {Gaudi}, {OGLE Collaboration}, {Jaroszy{\'n}ski}, {Kubiak},
  {Szyma{\'n}ski}, {Pietrzy{\'n}ski}, {Soszy{\'n}ski}, {Szewczyk}, {Ulaczyk},
  {Wyrzykowski}, {{\ensuremath{\mu}}FUN Collaboration}, {DePoy}, {Fox},
  {Gal-Yam}, {Han}, {L{\'e}pine}, {McCormick}, {Ofek}, {Park}, {Pogge}, {MOA
  Collaboration}, {Abe}, {Bennett}, {Bond}, {Britton}, {Gilmore}, {Hearnshaw},
  {Itow}, {Kamiya}, {Kilmartin}, {Korpela}, {Masuda}, {Matsubara}, {Motomura},
  {Muraki}, {Nakamura}, {Ohnishi}, {Okada}, {Rattenbury}, {Saito}, {Sako},
  {Sasaki}, {Sullivan}, {Sumi}, {Tristram}, {Yanagisawa}, {Yock}, {Yoshoika},
  {PLANET/RoboNet Collaborations}, {Albrow}, {Beaulieu}, {Brillant}, {Calitz},
  {Cassan}, {Cook}, {Coutures}, {Dieters}, {Dominis Prester}, {Donatowicz},
  {Fouqu{\'e}}, {Greenhill}, {Hill}, {Hoffman}, {Horne}, {J{\o}rgensen},
  {Kane}, {Kubas}, {Marquette}, {Martin}, {Meintjes}, {Menzies}, {Pollard},
  {Sahu}, {Vinter}, {Wambsganss}, {Williams}, {Bode}, {Bramich}, {Burgdorf},
  {Snodgrass}, {Steele}, {Doublier}, \& {Foellmi}}]{Dong2009}
{Dong}, S., {Gould}, A., {Udalski}, A., {et~al.} 2009,
  \href{http://dx.doi.org/10.1088/0004-637X/695/2/970}{\JournalTitle{\apj},
  695, 970}

\bibitem[{{Fukui} {et~al.}(2015){Fukui}, {Gould}, {Sumi}, {Bennett}, {Bond},
  {Han}, {Suzuki}, {Beaulieu}, {Batista}, {Udalski}, {Street}, {Tsapras},
  {Hundertmark}, {Abe}, {Bhattacharya}, {Freeman}, {Itow}, {Ling}, {Koshimoto},
  {Masuda}, {Matsubara}, {Muraki}, {Ohnishi}, {Philpott}, {Rattenbury},
  {Saito}, {Sullivan}, {Tristram}, {Yonehara}, {MOA Collaboration}, {Choi},
  {Christie}, {DePoy}, {Dong}, {Drummond}, {Gaudi}, {Hwang}, {Kavka}, {Lee},
  {McCormick}, {Natusch}, {Ngan}, {Park}, {Pogge}, {Shin}, {Tan}, {Yee},
  {{\ensuremath{\mu}}FUN Collaboration}, {Szyma{\'n}ski}, {Pietrzy{\'n}ski},
  {Soszy{\'n}ski}, {Poleski}, {Koz{\l}owski}, {Pietrukowicz}, {Ulaczyk},
  {Wyrzykowski}, {OGLE Collaboration}, {Bramich}, {Browne}, {Dominik}, {Horne},
  {Ipatov}, {Kains}, {Snodgrass}, {Steele}, \& {RoboNet
  Collaboration}}]{Fukui2015}
{Fukui}, A., {Gould}, A., {Sumi}, T., {et~al.} 2015,
  \href{http://dx.doi.org/10.1088/0004-637X/809/1/74}{\JournalTitle{\apj}, 809,
  74}

\bibitem[{{Furusawa} {et~al.}(2013){Furusawa}, {Udalski}, {Sumi}, {Bennett},
  {Bond}, {Gould}, {J{\o}rgensen}, {Snodgrass}, {Dominis Prester}, {Albrow},
  {Abe}, {Botzler}, {Chote}, {Freeman}, {Fukui}, {Harris}, {Itow}, {Ling},
  {Masuda}, {Matsubara}, {Miyake}, {Muraki}, {Ohnishi}, {Rattenbury}, {Saito},
  {Sullivan}, {Suzuki}, {Sweatman}, {Tristram}, {Wada}, {Yock}, {MOA
  Collaboration}, {Szyma{\'n}ski}, {Soszy{\'n}ski}, {Kubiak}, {Poleski},
  {Ulaczyk}, {Pietrzy{\'n}ski}, {Wyrzykowski}, {OGLE Collaboration}, {Choi},
  {Christie}, {DePoy}, {Dong}, {Drummond}, {Gaudi}, {Han}, {Hung}, {Hwang},
  {Lee}, {McCormick}, {Moorhouse}, {Natusch}, {Nola}, {Ofek}, {Pogge}, {Shin},
  {Skowron}, {Thornley}, {Yee}, {{\ensuremath{\mu}}FUN Collaboration},
  {Alsubai}, {Bozza}, {Browne}, {Burgdorf}, {Calchi Novati}, {Dodds},
  {Dominik}, {Finet}, {Gerner}, {Hardis}, {Harps{\o}e}, {Hinse}, {Hundertmark},
  {Kains}, {Kerins}, {Liebig}, {Mancini}, {Mathiasen}, {Penny}, {Proft},
  {Rahvar}, {Ricci}, {Scarpetta}, {Sch{\"a}fer}, {Sch{\"o}nebeck},
  {Southworth}, {Surdej}, {Wambsganss}, {MiNDSTEp Consortium}, {Street},
  {Bramich}, {Steele}, {Tsapras}, {RoboNet Collaboration}, {Horne},
  {Donatowicz}, {Sahu}, {Bachelet}, {Batista}, {Beatty}, {Beaulieu}, {Bennett},
  {Black}, {Bowens-Rubin}, {Brillant}, {Caldwell}, {Cassan}, {Cole},
  {Corrales}, {Coutures}, {Dieters}, {Fouqu{\'e}}, {Greenhill}, {Henderson},
  {Kubas}, {Marquette}, {Martin}, {Menzies}, {Shappee}, {Williams}, {Wouters},
  {van Saders}, {Zellem}, {Zub}, \& {PLANET Collaboration}}]{Furusawa2013}
{Furusawa}, K., {Udalski}, A., {Sumi}, T., {et~al.} 2013,
  \href{http://dx.doi.org/10.1088/0004-637X/779/2/91}{\JournalTitle{\apj}, 779,
  91}

\bibitem[{{Gaudi}(2012)}]{Gaudi2012}
{Gaudi}, B.~S. 2012,
  \href{http://dx.doi.org/10.1146/annurev-astro-081811-125518}{\JournalTitle{\araa},
  50, 411}

\bibitem[{{Gould}(2000)}]{Gould2000}
{Gould}, A. 2000, \href{http://dx.doi.org/10.1086/317037}{\JournalTitle{ApJ},
  542, 785}

\bibitem[{{Gould} {et~al.}(2006){Gould}, {Udalski}, {An}, {Bennett}, {Zhou},
  {Dong}, {Rattenbury}, {Gaudi}, {Yock}, {Bond}, {Christie}, {Horne},
  {Anderson}, {Stanek}, {DePoy}, {Han}, {McCormick}, {Park}, {Pogge},
  {Poindexter}, {Soszy{\'n}ski}, {Szyma{\'n}ski}, {Kubiak}, {Pietrzy{\'n}ski},
  {Szewczyk}, {Wyrzykowski}, {Ulaczyk}, {Paczy{\'n}ski}, {Bramich},
  {Snodgrass}, {Steele}, {Burgdorf}, {Bode}, {Botzler}, {Mao}, \&
  {Swaving}}]{Gould2006}
{Gould}, A., {Udalski}, A., {An}, D., {et~al.} 2006,
  \href{http://dx.doi.org/10.1086/505421}{\JournalTitle{\apjl}, 644, L37}

\bibitem[{Henry {et~al.}(1999)Henry, Franz, Wasserman, Benedict, Shelus, Ianna,
  Kirkpatrick, \& Donald W.~McCarthy}]{Henry1999}
Henry, T.~J., Franz, O.~G., Wasserman, L.~H., {et~al.} 1999,
  \href{http://dx.doi.org/10.1086/306793}{\JournalTitle{ApJ}, 512, 864}

\bibitem[{{Henry} \& {McCarthy}(1993)}]{Henry1993}
{Henry}, T.~J., \& {McCarthy}, Donald~W., J. 1993,
  \href{http://dx.doi.org/10.1086/116685}{\JournalTitle{AJ}, 106, 773}

\bibitem[{{Ida} \& {Lin}(2004)}]{Ida2004}
{Ida}, S., \& {Lin}, D.~N.~C. 2004,
  \href{http://dx.doi.org/10.1086/381724}{\JournalTitle{\apj}, 604, 388}

\bibitem[{{Johnson} {et~al.}(2020){Johnson}, {Penny}, {Gaudi}, {Kerins},
  {Rattenbury}, {Robin}, {Calchi Novati}, \& {Henderson}}]{Johnson2020}
{Johnson}, S.~A., {Penny}, M., {Gaudi}, B.~S., {et~al.} 2020,
  \href{http://dx.doi.org/10.3847/1538-3881/aba75b}{\JournalTitle{AJ}, 160,
  123}

\bibitem[{{Kennedy} {et~al.}(2006){Kennedy}, {Kenyon}, \&
  {Bromley}}]{Kennedy2006}
{Kennedy}, G.~M., {Kenyon}, S.~J., \& {Bromley}, B.~C. 2006,
  \href{http://dx.doi.org/10.1086/508882}{\JournalTitle{\apj}, 650, L139}

\bibitem[{{Koshimoto} {et~al.}(2021){Koshimoto}, {Baba}, \&
  {Bennett}}]{Koshimoto2021a}
{Koshimoto}, N., {Baba}, J., \& {Bennett}, D.~P. 2021,
  \href{http://dx.doi.org/10.3847/1538-4357/ac07a8}{\JournalTitle{\apj}, 917,
  78}

\bibitem[{{Koshimoto} {et~al.}(2020){Koshimoto}, {Bennett}, \&
  {Suzuki}}]{Naoki2020}
{Koshimoto}, N., {Bennett}, D.~P., \& {Suzuki}, D. 2020,
  \href{http://dx.doi.org/10.3847/1538-3881/ab8adf}{\JournalTitle{\aj}, 159,
  268}

\bibitem[{{Lissauer}(1993)}]{Lissauer1993}
{Lissauer}, J.~J. 1993,
  \href{http://dx.doi.org/10.1146/annurev.aa.31.090193.001021}{\JournalTitle{Annual
  Review of Astronomy and Astrophysics}, 31, 129}

\bibitem[{Lockhart {et~al.}(2019)Lockhart, Do, Larkin, Boehle, Campbell,
  Chappell, Chu, Ciurlo, Cosens, Fitzgerald, Ghez, Lu, Lyke, Mieda, Rudy,
  Vayner, Walth, \& Wright}]{Lockhart_2019}
Lockhart, K.~E., Do, T., Larkin, J.~E., {et~al.} 2019,
  \href{http://dx.doi.org/10.3847/1538-3881/aaf64e}{\JournalTitle{The
  Astronomical Journal}, 157, 75}

\bibitem[{{Lyke} {et~al.}(2017){Lyke}, {Do}, {Boehle}, {Campbell}, {Chappell},
  {Fitzgerald}, {Gasawy}, {Iserlohe}, {Krabbe}, {Larkin}, {Lockhart}, {Lu},
  {Mieda}, {McElwain}, {Perrin}, {Rudy}, {Sitarski}, {Vayner}, {Walth},
  {Weiss}, {Wizanski}, \& {Wright}}]{Lyke2017}
{Lyke}, J., {Do}, T., {Boehle}, A., {et~al.} 2017, {OSIRIS Toolbox:
  OH-Suppressing InfraRed Imaging Spectrograph pipeline}, Astrophysics Source
  Code Library, record ascl:1710.021,
  \href{http://arxiv.org/abs/1710.021}{{\sffamily ascl:1710.021}}

\bibitem[{{Minniti} {et~al.}(2010){Minniti}, {Lucas}, {Emerson}, {Saito},
  {Hempel}, {Pietrukowicz}, {Ahumada}, {Alonso}, {Alonso-Garcia}, {Arias},
  {Bandyopadhyay}, {Barb{\'a}}, {Barbuy}, {Bedin}, {Bica}, {Borissova},
  {Bronfman}, {Carraro}, {Catelan}, {Clari{\'a}}, {Cross}, {de Grijs},
  {D{\'e}k{\'a}ny}, {Drew}, {Fari{\~n}a}, {Feinstein}, {Fern{\'a}ndez
  Laj{\'u}s}, {Gamen}, {Geisler}, {Gieren}, {Goldman}, {Gonzalez}, {Gunthardt},
  {Gurovich}, {Hambly}, {Irwin}, {Ivanov}, {Jord{\'a}n}, {Kerins}, {Kinemuchi},
  {Kurtev}, {L{\'o}pez-Corredoira}, {Maccarone}, {Masetti}, {Merlo},
  {Messineo}, {Mirabel}, {Monaco}, {Morelli}, {Padilla}, {Palma}, {Parisi},
  {Pignata}, {Rejkuba}, {Roman-Lopes}, {Sale}, {Schreiber}, {Schr{\"o}der},
  {Smith}, {}, {Soto}, {Tamura}, {Tappert}, {Thompson}, {Toledo}, {Zoccali}, \&
  {Pietrzynski}}]{Minniti2010}
{Minniti}, D., {Lucas}, P.~W., {Emerson}, J.~P., {et~al.} 2010,
  \href{http://dx.doi.org/10.1016/j.newast.2009.12.002}{\JournalTitle{\na}, 15,
  433}

\bibitem[{{Nishiyama} {et~al.}(2009){Nishiyama}, {Tamura}, {Hatano}, {Kato},
  {Tanab{\'e}}, {Sugitani}, \& {Nagata}}]{Nishiyama2006}
{Nishiyama}, S., {Tamura}, M., {Hatano}, H., {et~al.} 2009,
  \href{http://dx.doi.org/10.1088/0004-637X/696/2/1407}{\JournalTitle{ApJ},
  696, 1407}

\bibitem[{{Penny} {et~al.}(2019){Penny}, {Gaudi}, {Kerins}, {Rattenbury},
  {Mao}, {Robin}, \& {Calchi Novati}}]{Penny2019}
{Penny}, M.~T., {Gaudi}, B.~S., {Kerins}, E., {et~al.} 2019,
  \href{http://dx.doi.org/10.3847/1538-4365/aafb69}{\JournalTitle{ApJs}, 241,
  3}

\bibitem[{{Poindexter} {et~al.}(2005){Poindexter}, {Afonso}, {Bennett},
  {Glicenstein}, {Gould}, {Szyma{\'n}ski}, \& {Udalski}}]{Pointdexter2005}
{Poindexter}, S., {Afonso}, C., {Bennett}, D.~P., {et~al.} 2005,
  \href{http://dx.doi.org/10.1086/468182}{\JournalTitle{\apj}, 633, 914}

\bibitem[{{Quenouille}(1949)}]{Quenouille1949}
{Quenouille}, M.~H. 1949,
  \href{http://dx.doi.org/10.1214/aoms/1177729989}{\JournalTitle{Ann. Math.
  Stat.}, 20, 355}

\bibitem[{{Quenouille}(1956)}]{Quenouille1956}
---. 1956,
  \href{http://dx.doi.org/10.1093/biomet/43.3-4.353}{\JournalTitle{Biometrika},
  43, 353}

\bibitem[{{Skowron} {et~al.}(2011){Skowron}, {Udalski}, {Gould}, {Dong},
  {Monard}, {Han}, {Nelson}, {McCormick}, {Moorhouse}, {Thornley}, {Maury},
  {Bramich}, {Greenhill}, {Koz{\l}owski}, {Bond}, {Poleski}, {Wyrzykowski},
  {Ulaczyk}, {Kubiak}, {Szyma{\'n}ski}, {Pietrzy{\'n}ski}, {Soszy{\'n}ski},
  {OGLE Collaboration}, {Gaudi}, {Yee}, {Hung}, {Pogge}, {DePoy}, {Lee},
  {Park}, {Allen}, {Mallia}, {Drummond}, {Bolt}, {{\ensuremath{\mu}}FUN
  Collaboration}, {Allan}, {Browne}, {Clay}, {Dominik}, {Fraser}, {Horne},
  {Kains}, {Mottram}, {Snodgrass}, {Steele}, {Street}, {Tsapras}, {RoboNet
  Collaboration}, {Abe}, {Bennett}, {Botzler}, {Douchin}, {Freeman}, {Fukui},
  {Furusawa}, {Hayashi}, {Hearnshaw}, {Hosaka}, {Itow}, {Kamiya}, {Kilmartin},
  {Korpela}, {Lin}, {Ling}, {Makita}, {Masuda}, {Matsubara}, {Muraki},
  {Nagayama}, {Miyake}, {Nishimoto}, {Ohnishi}, {Perrott}, {Rattenbury},
  {Saito}, {Skuljan}, {Sullivan}, {Sumi}, {Suzuki}, {Sweatman}, {Tristram},
  {Wada}, {Yock}, {MOA Collaboration}, {Beaulieu}, {Fouqu{\'e}}, {Albrow},
  {Batista}, {Brillant}, {Caldwell}, {Cassan}, {Cole}, {Cook}, {Coutures},
  {Dieters}, {Dominis Prester}, {Donatowicz}, {Kane}, {Kubas}, {Marquette},
  {Martin}, {Menzies}, {Sahu}, {Wambsganss}, {Williams}, {Zub}, \& {PLANET
  Collaboration}}]{Skowron2011}
{Skowron}, J., {Udalski}, A., {Gould}, A., {et~al.} 2011,
  \href{http://dx.doi.org/10.1088/0004-637X/738/1/87}{\JournalTitle{\apj}, 738,
  87}

\bibitem[{{Smith} {et~al.}(2003){Smith}, {Mao}, \& {Paczy{\'n}ski}}]{Smith2003}
{Smith}, M.~C., {Mao}, S., \& {Paczy{\'n}ski}, B. 2003,
  \href{http://dx.doi.org/10.1046/j.1365-8711.2003.06183.x}{\JournalTitle{\mnras},
  339, 925}

\bibitem[{{Spergel} {et~al.}(2015){Spergel}, {Gehrels}, {Baltay}, {Bennett},
  {Breckinridge}, {Donahue}, {Dressler}, {Gaudi}, {Greene}, {Guyon}, {Hirata},
  {Kalirai}, {Kasdin}, {Macintosh}, {Moos}, {Perlmutter}, {Postman},
  {Rauscher}, {Rhodes}, {Wang}, {Weinberg}, {Benford}, {Hudson}, {Jeong},
  {Mellier}, {Traub}, {Yamada}, {Capak}, {Colbert}, {Masters}, {Penny},
  {Savransky}, {Stern}, {Zimmerman}, {Barry}, {Bartusek}, {Carpenter}, {Cheng},
  {Content}, {Dekens}, {Demers}, {Grady}, {Jackson}, {Kuan}, {Kruk}, {Melton},
  {Nemati}, {Parvin}, {Poberezhskiy}, {Peddie}, {Ruffa}, {Wallace}, {Whipple},
  {Wollack}, \& {Zhao}}]{Spergel2015}
{Spergel}, D., {Gehrels}, N., {Baltay}, C., {et~al.} 2015, \JournalTitle{arXiv
  e-prints}, arXiv:1503.03757

\bibitem[{{Stetson}(1987)}]{Stetson1987}
{Stetson}, P.~B. 1987,
  \href{http://dx.doi.org/10.1086/131977}{\JournalTitle{\pasp}, 99, 191}

\bibitem[{{Surot} {et~al.}(2020){Surot}, {Valenti}, {Gonzalez}, {Zoccali},
  {S{\"o}kmen}, {Hidalgo}, \& {Minniti}}]{Surot2020}
{Surot}, F., {Valenti}, E., {Gonzalez}, O.~A., {et~al.} 2020,
  \href{http://dx.doi.org/10.1051/0004-6361/202038346}{\JournalTitle{AAP}, 644,
  A140}

\bibitem[{{Suzuki} {et~al.}(2016){Suzuki}, {Bennett}, {Sumi}, {Bond}, {Rogers},
  {Abe}, {Asakura}, {Bhattacharya}, {Donachie}, {Freeman}, {Fukui}, {Hirao},
  {Itow}, {Koshimoto}, {Li}, {Ling}, {Masuda}, {Matsubara}, {Muraki},
  {Nagakane}, {Onishi}, {Oyokawa}, {Rattenbury}, {Saito}, {Sharan}, {Shibai},
  {Sullivan}, {Tristram}, {Yonehara}, \& {MOA Collaboration}}]{Suzuki2016}
{Suzuki}, D., {Bennett}, D.~P., {Sumi}, T., {et~al.} 2016,
  \href{http://dx.doi.org/10.3847/1538-4357/833/2/145}{\JournalTitle{\apj},
  833, 145}

\bibitem[{{Szyma{\'n}ski} {et~al.}(2011){Szyma{\'n}ski}, {Udalski},
  {Soszy{\'n}ski}, {Kubiak}, {Pietrzy{\'n}ski}, {Poleski}, {Wyrzykowski}, \&
  {Ulaczyk}}]{Szymanski2011}
{Szyma{\'n}ski}, M.~K., {Udalski}, A., {Soszy{\'n}ski}, I., {et~al.} 2011,
  \href{http://dx.doi.org/10.48550/arXiv.1107.4008}{\JournalTitle{\actaa}, 61,
  83}

\bibitem[{{Terry} {et~al.}(2021){Terry}, {Bhattacharya}, {Bennett}, {Beaulieu},
  {Koshimoto}, {Blackman}, {Bond}, {Cole}, {Henderson}, {Lu}, {Marquette},
  {Ranc}, \& {Vandorou}}]{Terrry2021}
{Terry}, S.~K., {Bhattacharya}, A., {Bennett}, D.~P., {et~al.} 2021,
  \href{http://dx.doi.org/10.3847/1538-3881/abcc60}{\JournalTitle{\aj}, 161,
  54}

\bibitem[{{Terry} {et~al.}(2022){Terry}, {Bennett}, {Bhattacharya},
  {Koshimoto}, {Beaulieu}, {Blackman}, {Bond}, {Cole}, {Lu}, {Marquette},
  {Ranc}, {Rektsini}, \& {Vandorou}}]{Terry2022}
{Terry}, S.~K., {Bennett}, D.~P., {Bhattacharya}, A., {et~al.} 2022,
  \href{http://dx.doi.org/10.3847/1538-3881/ac9518}{\JournalTitle{\aj}, 164,
  217}

\bibitem[{{Terry} {et~al.}(2024){Terry}, {Beaulieu}, {Bennett}, {Hamdorf},
  {Bhattacharya}, {Chaudhry}, {Cole}, {Koshimoto}, {Anderson}, {Bachelet},
  {Blackman}, {Bond}, {Lu}, {Marquette}, {Ranc}, {Rektsini}, {Sahu}, \&
  {Vandorou}}]{Terry2024}
{Terry}, S.~K., {Beaulieu}, J.-P., {Bennett}, D.~P., {et~al.} 2024,
  \href{http://dx.doi.org/10.3847/1538-3881/ad5444}{\JournalTitle{\aj}, 168,
  72}

\bibitem[{{Tsapras} {et~al.}(2009){Tsapras}, {Street}, {Horne}, {Snodgrass},
  {Dominik}, {Allan}, {Steele}, {Bramich}, {Saunders}, {Rattenbury}, {Mottram},
  {Fraser}, {Clay}, {Burgdorf}, {Bode}, {Lister}, {Hawkins}, {Beaulieu},
  {Fouqu{\'e}}, {Albrow}, {Menzies}, {Cassan}, \&
  {Dominis-Prester}}]{Tsapras2009}
{Tsapras}, Y., {Street}, R., {Horne}, K., {et~al.} 2009,
  \href{http://dx.doi.org/10.1002/asna.200811130}{\JournalTitle{Astronomische
  Nachrichten}, 330, 4}

\bibitem[{Tukey(1958)}]{Tukey1958}
Tukey, J.~W. 1958,
  \href{http://dx.doi.org/10.1214/aoms/1177706635}{\JournalTitle{The Annals of
  Mathematical Statistics}, 29, 581 }

\bibitem[{{Udalski} {et~al.}(2005){Udalski}, {Jaroszy{\'n}ski},
  {Paczy{\'n}ski}, {Kubiak}, {Szyma{\'n}ski}, {Soszy{\'n}ski},
  {Pietrzy{\'n}ski}, {Ulaczyk}, {Szewczyk}, {Wyrzykowski}, {OGLE
  Collaboration}, {Christie}, {DePoy}, {Dong}, {Gal-Yam}, {Gaudi}, {Gould},
  {Han}, {L{\'e}pine}, {McCormick}, {Park}, {Pogge}, {{\ensuremath{\mu}}FUN
  Collaboration}, {Bennett}, {Bond}, {Muraki}, {Tristram}, {Yock}, {MOA
  Collaboration}, {Beaulieu}, {Bramich}, {Dieters}, {Greenhill}, {Hill},
  {Horne}, {Kubas}, \& {PLANET/ROBONET Collaboration}}]{Udalski2005}
{Udalski}, A., {Jaroszy{\'n}ski}, M., {Paczy{\'n}ski}, B., {et~al.} 2005,
  \JournalTitle{\apjl}

\bibitem[{{Vandorou} {et~al.}(2020){Vandorou}, {Bennett}, {Beaulieu}, {Alard},
  {Blackman}, {Cole}, {Bhattacharya}, {Bond}, {Koshimoto}, \&
  {Marquette}}]{Vandorou2020}
{Vandorou}, A., {Bennett}, D.~P., {Beaulieu}, J.-P., {et~al.} 2020,
  \href{http://dx.doi.org/10.3847/1538-3881/aba2d3}{\JournalTitle{\aj}, 160,
  121}

\end{thebibliography}

\end{document}